\documentclass{emulateapj}
\pdfoutput=1
\usepackage{graphicx}
\usepackage{subfigure}
\usepackage{natbib}
\usepackage{ifpdf}

\shorttitle{Structure and Dynamics of WLM}
\shortauthors{Leaman et al.}

\begin{document}

\title{The Resolved Structure and Dynamics of an Isolated Dwarf Galaxy: A VLT and Keck Spectroscopic Survey of WLM}
\author{Ryan Leaman$^{1}$, Kim A. Venn$^{1}$, Alyson M. Brooks$^{2}$, Giuseppina Battaglia$^{3}$, Andrew A. Cole$^{4}$, Rodrigo A. Ibata$^{5}$, Mike J. Irwin$^{6}$, Alan W. McConnachie$^{7}$, J. Trevor Mendel$^{1}$, Eline Tolstoy$^{8}$}
\affil{$^{1}$University of Victoria, Canada, $^{2}$California Institute of Technology, US, $^{3}$European Southern Observatory, Germany, $^{4}$University of Tasmania, Australia, $^{5}$Observatoire de Strasbourg, France, $^{6}$Institute of Astronomy, University of Cambridge, UK $^{7}$National Research Council of Canada, Herzberg Institute of Astrophysics, Canada $^{8}$Kapteyn Astronomical Institute, University of Groningen, The Netherlands}
\email{rleaman@uvic.ca}

\begin{abstract}
We present spectroscopic data for 180 red giant branch stars in the isolated dwarf irregular galaxy WLM.  Observations of the Calcium II triplet lines in spectra of RGB stars covering the entire galaxy were obtained with FORS2 at the VLT and DEIMOS on Keck II allowing us to derive velocities, metallicities, and ages for the stars. With accompanying photometric and radio data we have measured the structural parameters of the stellar and gaseous populations over the full galaxy.  The stellar populations show an intrinsically thick configuration with $0.39 \leq q_{0} \leq 0.57$.  The stellar rotation in WLM is measured to be $17 \pm 1$ km s$^{-1}$, however the ratio of rotation to pressure support for the stars is $V/\sigma \sim 1$, in contrast to the gas whose ratio is seven times larger.  This, along with the structural data and alignment of the kinematic and photometric axes, suggests we are viewing WLM as a highly inclined oblate spheroid.  Stellar rotation curves, corrected for asymmetric drift, are used to compute a dynamical mass of $4.3\pm 0.3\times10^{8} $M$_{\odot}$ at the half light radius ($r_{h} = 1656 \pm 49$ pc).  The stellar velocity dispersion increases with stellar age in a manner consistent with giant molecular cloud and substructure interactions producing the heating in WLM.  Coupled with WLM's isolation, this suggests that the extended vertical structure of its stellar and gaseous components and increase in stellar velocity dispersion with age are due to internal feedback, rather than tidally driven evolution.  These represent some of the first observational results from an isolated Local Group dwarf galaxy which can offer important constraints on how strongly internal feedback and secular processes modulate SF and dynamical evolution in low mass isolated objects.
\end{abstract}


\section{Introduction}
In models of hierarchical structure formation such as the current $\rm{\Lambda}$CDM cosmologies, dwarf galaxies are fundamental components of the mass assembly history of larger galaxies, such as our own Milky Way (MW).  With dynamical masses below $10^{10}$ M$_{\odot}$ these objects offer important observational tests for models, as they represent the smallest groupings of baryons out of which galaxies may be built (e.g., \citealt{NFW97, Moore99, Madau01}).  Analyzing the survival of these low mass objects, particularly through reionisation, is important to constraining galaxy formation models \citep{Ricotti05, Gnedin06}.  Understanding the factors that allow low mass dwarf galaxies to survive to the present day may help explain discrepancies between observed and predicted distribution of subhalos around the MW.  This requires an understanding of how internal and environmental effects are expected to shape the evolution of low mass systems like dwarf galaxies.

With much shallower potential wells and lower metallicities than their higher mass counterparts, dwarf galaxies offer crucial laboratories in understanding how star formation (SF) proceeds throughout the lifetime of low mass objects. With total masses in the range $10^{7} - 10^{10}$ M$_{\odot}$, dwarf galaxies have been shown to exhibit star formation efficiencies much lower than higher mass galaxies \citep{Roychowdhury09}.  

Environmental feedback offers one possibility for influencing the stellar mass assembly rate.  For example, is there a minimum halo mass which determines a galaxy's likelihood of retaining baryons through reionisation (e.g., \citealt{NFW97, Madau01, Ricotti05, Bovill10,Sawala11})? The apparent morphology-density relation in the Local Group, whereby more gas rich dwarf irregulars (dIrrs) are found at larger distances than the closer gas poor dwarf spheroidals (dSphs) \citep{Einasto74, vdb99}, is often invoked as evidence that environmental feedback has played an active role in low mass dwarfs.  In this case both tidal and ram pressure stripping by the Milky Way could drastically alter the gas content and structure of dwarf galaxies, certainly playing a role in how their stellar content evolves over a cosmic time (e.g., \citealt{Mayer01b, Kazantzidis11, Lokas10}).

Internal feedback such as SF and supernova (SN) driven winds may also play a role in modulating the stellar mass buildup of dwarf galaxies.  In that case, due to the low potential well of the dwarf galaxies, the gas may not be retained during large episodes of star formation, shortening the SF lifetime in some cases \citep{DekelWoo03}.  This has been shown to have effects on not just the chemical enrichment history, but also the structure of low mass dwarfs \citep{Stinson09,Sotnikova03,Governato10}.  With or without gas loss internal heating may regulate the SF efficiency, leading to lower star formation rates \citep{Brooks07} and thicker structure \citep{Kaufmann07}.  Alternatively, low star formation efficiencies in dwarf galaxies may be due to the dependency of SF on H$_{2}$.  Recent work \citep{Bigiel08,Leroy08,Schruba11} indicates that SF density in a galaxy is most directly correlated with cold molecular hydrogen, H$_{2}$, rather than the total gas surface density.  Lower mass systems with low gas column densities do not allow self shielding of the H$_{2}$, resulting in dissociation by the local radiation field \citep{Robertson08, Kuhlen11}. 

 Further suppression of H$_{2}$ formation is expected for low ISM metallicites, where fewer sites for dust formation and longer cooling times for the gas would be present \citep{KrumDekel11}.  If the SF efficiencies in dwarf galaxies are lowered due to their intrinsic low mass and metal content (both of which hinder the gas reaching the molecular phase), the amount of feedback need not be as large in order to explain the extended low level of SF \citep{Kuhlen11}.  However the constraints on what relative contribution feedback still must play is not well known currently.  Establishing whether low star formation rates in dwarf galaxies are due to feedback or H$_{2}$ regulated mechanisms can be answered in part through analysis of nearby dwarf galaxies. 
 
Mechanisms that shape the evolution of the dwarf galaxy, whether internal or external, may impart observational signatures onto the stellar and gaseous populations in the galaxies.  For example, the amount of pressure versus rotational support is often tracked in simulations of dwarf galaxy evolution \citep{Kazantzidis11}.  In the simulations of \cite{Mayer01b, Kazantzidis11, Lokas10}, dIrrs with thin cold gas disks ($2 \leq V/\sigma \leq 5$) are tidally perturbed upon pericentre passage of the MW - a process invoked to explain the gas loss and change in stellar orbits that would be necessary to transform them into dSphs ($V/\sigma \lesssim 0.5$).  However alternative theories exist in which the progenitors of dSphs form in hotter, thicker disks or spheroids \citep{Kaufmann07} - this may in turn influence the SF efficiency as the molecular gas formation rate would be influenced by the disk morphology \citep{Robertson08}.  And recent theoretical \citep{Governato10} and observational work \citep{SJ10,Roychowdhury10}, suggest that significant rearrangement and evolution of baryonic structure can occur through secular evolution - whether via feedback (SF, SNe), stellar migration, or global disk instabilities \citep{Stinson09,Sotnikova03}.  
However, to connect the enrichment and kinematic history of the galaxy over most of the age of the Universe (the last $\sim 12$ Gyrs), spectroscopic data (of resolved RGB stars) is required.  In particular the relative amount of rotation and velocity dispersion of the stellar populations will evolve over the lifetime of the dwarf galaxy, and can be traced as a function of time.  These spectroscopic observations characterize the changing dynamics of the tracer populations over the lifetime of the dwarf (e.g., \citealt{vdM06}).  This can be combined with deep photometric imaging studies, which trace the \emph{current} structure of the galaxy \citep{IH95}.  Additionally stellar density profiles yield an estimate of the baryonic concentration, crucial for understanding the mass profiles \citep{vandenbosch01}.  Finally, coarse metallicity information on a global scale can reveal chemical gradients across the galaxy.  

Therefore with structural analyses from photometric data, and the kinematics and ages from spectroscopic observations, it is possible to build a picture of how the mass distribution and dynamics of the galaxy evolved.   While both internal and environmental feedback may have clear signatures on the dynamics of stellar and gaseous populations in low mass dwarfs, the question remains to what degree one effect dominates for a given dwarf galaxy.  In isolated dwarf galaxies any changes are more likely due to internal effects - offering a window into how effective secular processes are at transforming a galaxy in the absence of strong tides.  Any limits which can be placed on how feedback operates in these remote dwarf galaxies will offer important constraints in simulations of SF in low mass dwarfs - giving an observational framework that can constrain models of internal feedback mechanisms.

Constraining the relative amount of internal feedback driven evolution requires studying galaxies that are likely to have been isolated over most of their history, unlike the nearby dSphs that have been influenced by the MW.  While dSph distances ($\lesssim 250$kpc) put them well within the reach of modern 8m class spectrographs and imagers, they have most likely been tidally influenced by the MW, complicating interpretation.  To date the large distances ($\geq 500$kpc) of isolated dIrrs have prevented substantive spectroscopic surveys, although these are the prime candidates to unravel the nature versus nurture question.  Stellar structure and dynamics of the populations in the LMC have been studied in the past (c.f., \citealt{vdM06}), but due to its larger mass, close distance to the MW, and interaction with the SMC it is not ideal for studies of internal feedback driven evolution in a low mass dwarf galaxy.  Ideally we wish to study a dwarf galaxy at large galactocentric distance, in which the evolved stars will likely have had fewer tidal distortions.  With this we can understand for the first time, the dynamical state of the stellar component in an isolated dIrr.  This can help constrain the initial conditions used for dwarf galaxy tidal transformation scenarios (i.e., \citealt{Mayer06}), which may be important for the relative timescales of those models \citep{Kazantzidis11}.  

Galaxies such as the Wolf-Lundmark-Melotte dIrr (WLM, DDO 221; \citealt{Wolf10,Melotte26}), sit at large galactocentric distances, and their complex stellar and gaseous populations trace how their structure and dynamics evolve over a cosmic lifetime, with minimal external environmental influence.  Observations of the bright supergiant population \citep{Venn04, Bresolin06, Urbaneja08} and ISM \citep{Skillman89,Hodge95, Lee05} sampled the young populations of WLM, but offered little insight into the earlier epochs of formation and evolution.  Similarly the photometric studies \citep{Ables77, Minniti97, Hodge99, Battinelli04, Alan05, Jackson07a} were only able to provide global views of the evolved population and were subject to degeneracies in age and metallicity.  We seek to answer two main questions on dwarf galaxy evolution: (1) what is the role of internal vs. external feedback in shaping the structural and dynamical properties of the dwarf? and (2) what modulates the star formation efficiencies in low metallicity low mass systems - SNe feedback or H$_{2}$ regulated SF modes?  To answer these, there needs to be a spectroscopic survey of RGB stars spanning 10 Gyrs in age, which produces abundances and velocities in an isolated dwarf galaxy like WLM.
In \cite{Leaman09} (hereafter Paper I), we presented spectra for 78 stars from FORS2 on VLT, which represented the first medium scale resolved spectroscopic study of an \emph{isolated} dwarf galaxy in the Local Group.  In this paper we more than double our sample size, allowing for the first time a chance to study these questions in a truly isolated dwarf galaxy.

\subsection{Quantifying WLM's Isolation}
It is useful at this point to discuss the environment that WLM occupies within the Local Group, as the goal of this paper is to minimize the relative importance of external (tides, ram pressure) effects in order to learn more about the possible impact of internal feedback effects.  

\begin{figure}
\begin{center}
\ifpdf
\includegraphics[width=0.9\textwidth]{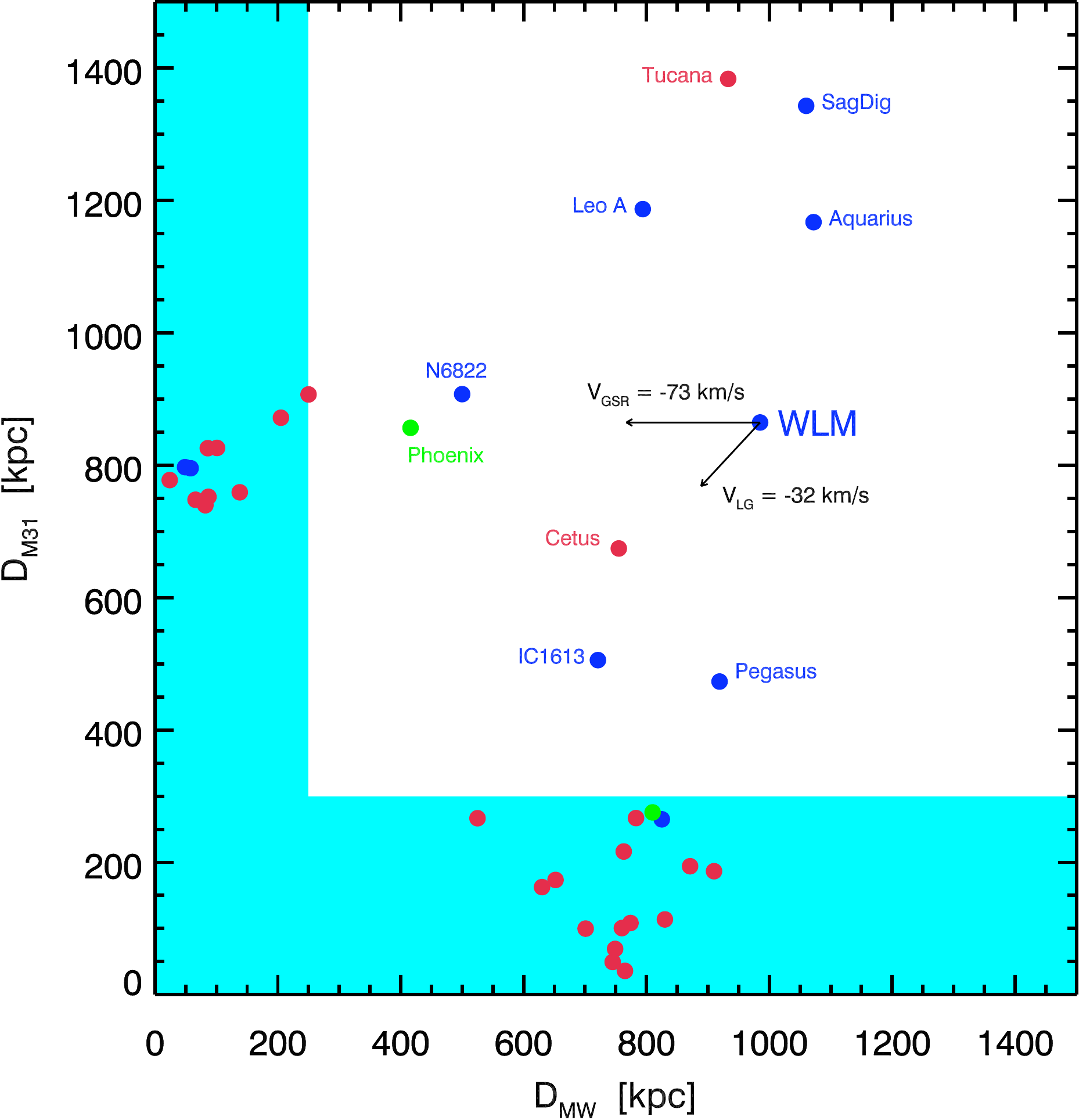}
\else
\includegraphics[angle=270,width=0.9\textwidth]{dmwdm31.eps}
\fi
\caption{Distance from M31 and the Milky Way for Local Group dwarf irregulars (\emph{blue}), dwarf spheroidals (\emph{red}), and transition dwarfs (\emph{green}).  Shown are the projected galactocentric standard of rest (GSR) and Local Group standard of rest velocities for WLM.  Evident is WLM's large isolation from the two massive spiral galaxies, as well as other dwarf galaxies. Coordinates are taken from \cite{Mateo98}.  Shaded areas correspond to the approximate virial radii of the Milky Way and M31.}
\label{fig:wlmiso}
\end{center}
\end{figure}

Derived and adopted properties for WLM are shown in Table 1 for clarity.  WLM lies approximately 1 Mpc \citep{Gorski11} from both the MW and M31, with its nearest neighbour being the small dSph Cetus ($M_{half} \sim 9 \times 10^{7} $M$_{\odot}$; \citealt{Walker09}), at 250 kpc away \citep{Whiting99}, shown in Figure \ref{fig:wlmiso}.  \cite{Karachentsev05} calculate a tidal index for WLM of $\Theta = 0.3$, where they define $\Theta \equiv {\rm max [log}(M_{k}/D_{ik}^{3}){\rm ] + C}$ to be the amount a galaxy is acted on by its largest tidal disturber\footnote{As noted in that work, $M_{k}$ is the $k$th disturber galaxy at a distance $D_{ik}$, and $\Theta = 0$ corresponds to an object with a Keplerian orbital period about MW/M31 equal to $1/H$.  For reference, Sgr dSph has $\Theta = 5.6$} . With the exception of Tucana (-0.2), Pegasus (-0.1), Aquarius (-0.1), and Leo A (0.2),  WLM is one of the five least tidally disturbed and most isolated galaxies within a $\sim 1$ Mpc sphere of the MW.  Projection of WLM's heliocentric velocity ($v_{sys} \sim -130$ km s$^{-1}$) towards the Local Group centre of mass results in a velocity with respect to the barycentre of $v_{LG} = -32$ km s$^{-1}$, implying that it has just passed apocentre.  With its current distance and that velocity, and an inferred Local Group mass of $5.6\times10^{12}$M$_{\odot}$ (based on M31-MW orbit timing arguments which assume an age of 13.7 Gyr for the universe; c.f., \citealt{Lyndenbell81}), the maximum apocentre that WLM could have is approximately 1.3 Mpc.  Assuming a completely radial orbit, the implied orbital period for WLM to reach its pericentre with the Milky Way is  $11-17$ Gyr.  Thus WLM, in addition to being currently quite isolated, has had at most one pericentre passage in its lifetime (which would have been at least 11 Gyr ago at $z \simeq 2.5$), meaning its total evolution has been much less dominated by tidal and ram pressure effects from the Milky Way than other Local Group galaxies.\\
\\
The paper will be presented as follows: in $\S 2$ we will present the data sets with a focus on the observations and reductions.  $\S 3$ will discuss the spectral analysis techniques used to reduce and analyze the data in a homogeneous way to extract the chemodynamical properties for this study.  $\S 4$ will present results on the structure and kinematics of the stellar and gaseous populations, $\S 5$ a discussion of the mass estimates, and $\S 6$ will focus on the evolution of the stellar dynamics.

\section{Observations and Data Reductions}
\begin{figure}
\begin{center}
\ifpdf
\includegraphics[width=0.9\textwidth]{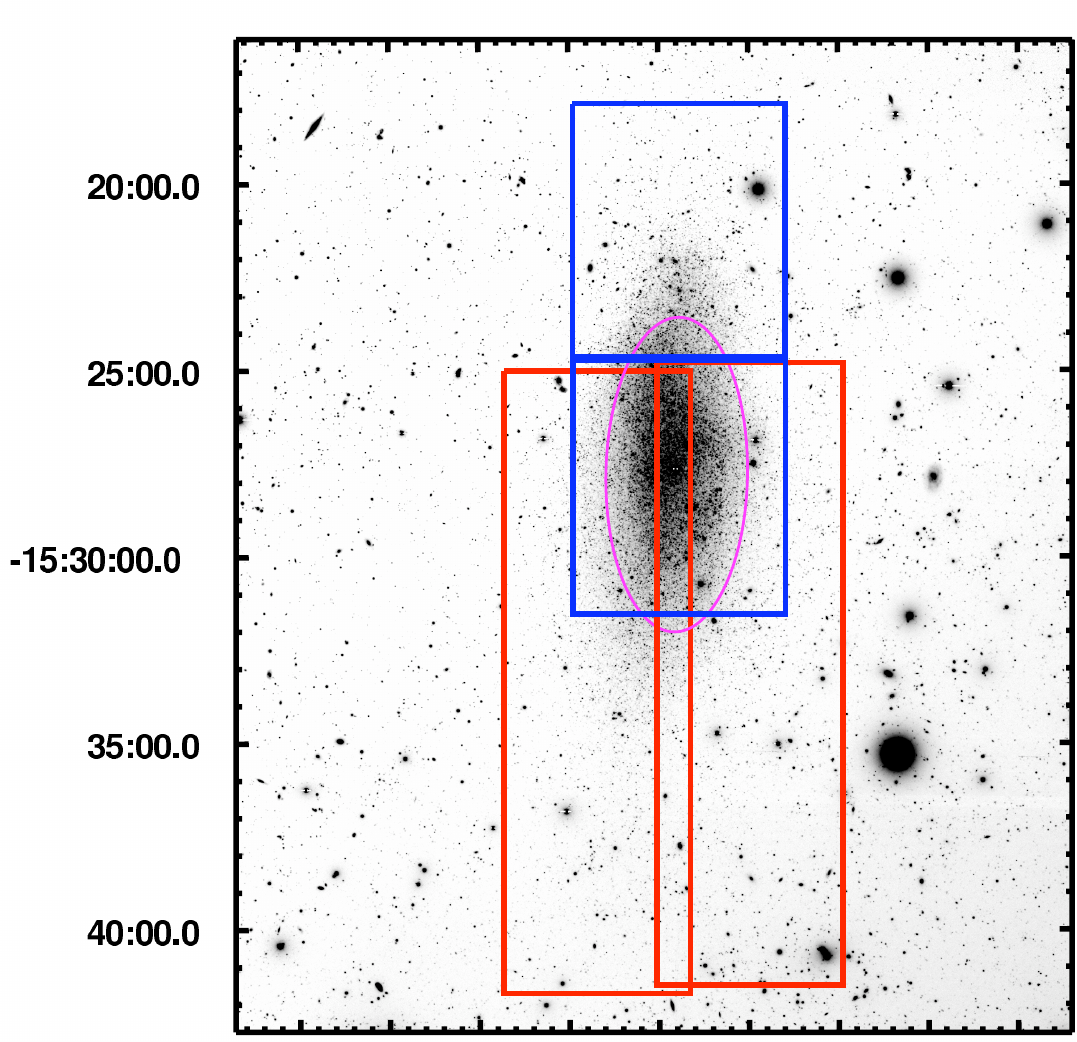}
\else
\includegraphics[width=0.9\textwidth]{wlmfov11g2.eps}
\vspace{5mm}
\fi
\caption{Portion of an I band image of WLM from the MOSAIC-II camera at the 4m Blanco Telescope at CTIO (Leaman et al., in prep.).  The total image is 
approximately $36'\times 36'$, equal to the coverage from our INT WFC photometry.  This Figure shows a zoomed in region of $22' \times 26'$, with North being up and East to the left.
The relative locations of the VLT FORS2 (\textit{blue}), and Keck II DEIMOS (\textit{red}) spectroscopic fields are shown. Magenta ellipse marks the half light radius assuming an ellipticity of $e = 0.55$ and photometric position angle of $179^{\circ}$.  The two A-type supergiants 
from \cite{Venn03} are located approximately in the centre of our lower FORS2 field, along with the HII regions from \cite{Hodge95}, and the B supergiants from \cite{Bresolin06}.  
}
\label{fig:fov}
\end{center}
\end{figure}

The data presented in this paper comes from observational campaigns on several instruments and telescopes.  78 stars were observed with the FORS2 \citep{FORS2} spectrograph on VLT and this data was presented in \cite{Leaman09}.  These spectroscopic observations are supplemented here with spectra of 140 new stars observed at higher resolution with the DEIMOS \citep{Faber03} spectrograph on Keck II .  V and i band photometry from the INT Wide Field Camera covering a $36^{'} \times 36^{'}$ field of view was used for several analysis steps.  This photometric data is presented in \cite{Alan05} and has been converted to V and I using the calibration in that work.  Near infrared JHK data from the UKIRT/WFCAM telescope was also used \citep{Tatton10}.  In addition we use VLA radio observations of WLM from \cite{Kepley07} which the authors kindly made available for use here (A. Kepley, priv. comm.).
Figure \ref{fig:fov} illustrates the coverage of our slitmasks for both the FORS2 and DEIMOS observations.

The new DEIMOS observations were taken during 3 nights of September of 2009.  Two $16^{'} \times 5^{'}$ slitmasks covering the body of WLM were observed (see Fig. \ref{fig:fov}) during seeing that ranged between $0.7^{''}-1.2^{''}$, with a median value of $0.8^{''}$.  The instrument setup used the 1200l/mm grating with OG550 blocking filter, which yielded a spectral resolution of $\sim 1.4 \rm{\AA}$ through $1.0^{''}$ slits.  The spectral range spanned roughly $7800 - 9300 \AA$.  Wavelength calibration was aided with the use of NeArKrXe arc lamp exposures, in addition to the standard, quartz flat field calibrating exposures.  Each mask had $10\times30$ minute exposures plus $3\times20$ minute exposures, all taken during good seeing and weather conditions, yielding 6 hours integration time per slit.\\

\begin{figure}
\begin{center}
\ifpdf
\includegraphics[width=0.98\textwidth]{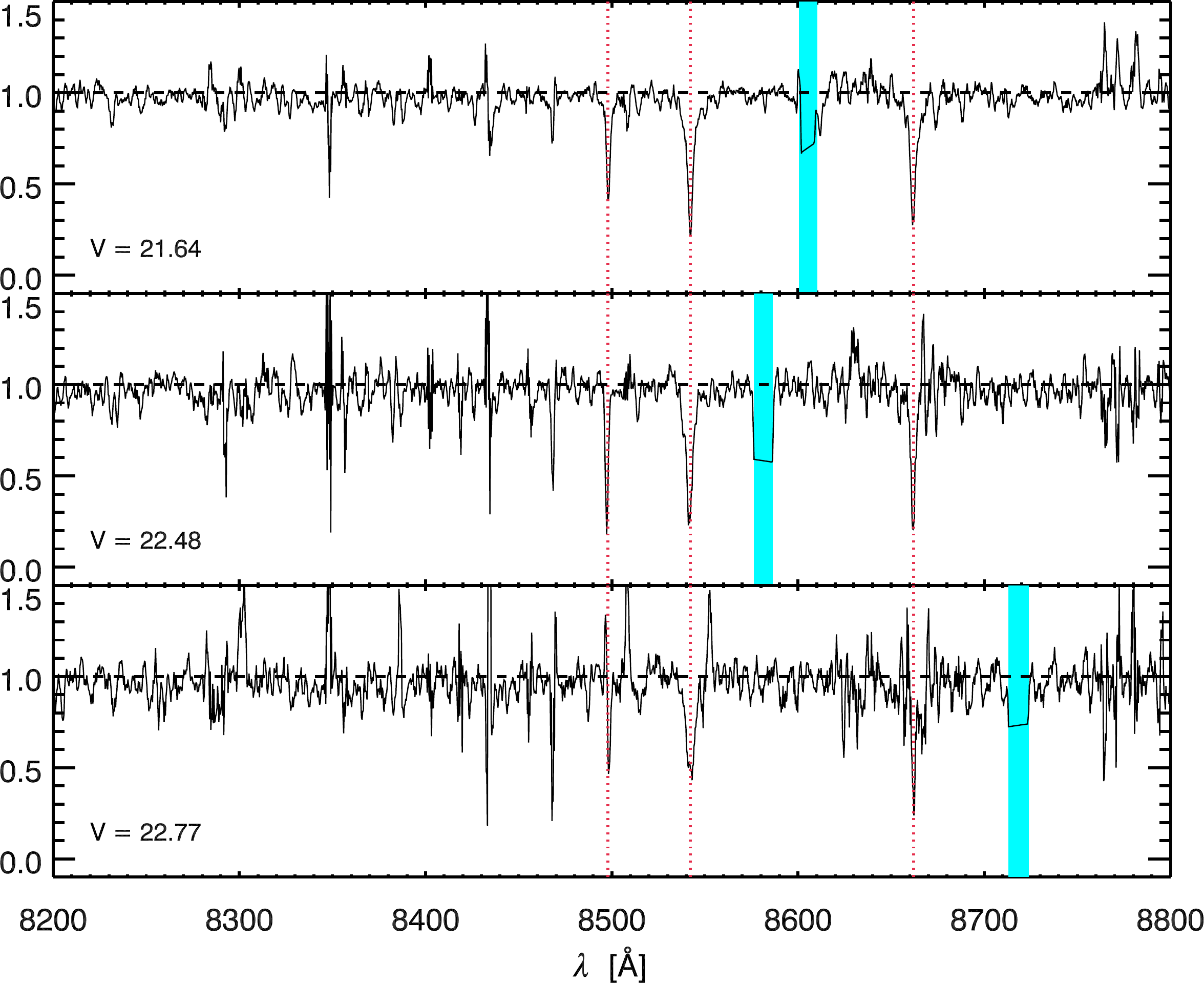}
\else
\includegraphics[angle=270,width=0.98\textwidth]{keckspexb2.eps}
\fi
\caption{Example DEIMOS co-added spectra for three member stars in our sample of varying magnitudes, plotted in arbitrary units of flux and continuum normalized and binned by a factor of 4. Visible are the three strong Ca II triplet lines at $\lambda \sim 8498, 8542, $ and 8662$\rm{\AA}$ shown by the red dotted lines, as well as the DEIMOS chip gap (blue regions).  The signal to noise per angstrom for the spectra are $\sim 30$, 19, and 15 respectively.}
\label{fig:keckspex}
\end{center}
\end{figure}

As the DEEP2 reduction pipeline is not optimized for faint stellar spectra, the data were pipeline processed using a privately written reduction procedure (R. Ibata).  In brief the pipeline performs the requisite calibration tasks to characterize the slit position on the CCD and remove any instrumental signatures, and non-orthogonal projection biases.  Details of the reduction procedure used for the DEIMOS spectra is described in \cite{Ibata11} and references therein.  Extraction of the spectra in each slit results in a single stellar source which is further subject to wavelength calibration and continuum normalized, before being radial velocity corrected via cross correlation against a suite of template spectra.  Due to the extreme faintness of the stars, even 10m class telescopes require coaddition of 6-8 hours of integration time to produce S/N $\geq 10 {\rm \AA^{-1}}$ data appropriate for resolved chemodynamic studies such as this.  Due to concerns about the poor performance of the flexure compensation system on DEIMOS at the time of our observations, the spectra were extracted individually and processed as single data products and coadded later.  The signal to noise per angstrom for the FORS2 spectra ranged from $17 \leq {\rm S/N} \leq 30$, and $6 \leq {\rm S/N} \leq 30$ for the DEIMOS data.  Figure \ref{fig:keckspex} shows three representative spectra of stars observed with our DEIMOS configuration.

In addition to the two masks on WLM three galactic clusters were observed for calibration purposes - the old open cluster NGC 6791, and globular clusters Pal 14 and NGC 7078.  Single mask setups yielded $5-50$ stars per cluster, which were observed 2-3 times per night throughout the observing run.  This allowed us to check the precision of our metallicities and assure that we are bringing the [Fe/H] estimates onto a common, calibrated scale.

\section{Spectral Analysis}
\subsection{Radial Velocity Measurements} 
Radial velocities were measured from the three strong calcium II triplet (CaT) lines ($\lambda \sim 8498, 8542, 8662 \rm{\AA}$) which 
in the case of the DEIMOS data, had a velocity correction applied from the sky 
OH lines in order to mitigate errors due to the instrument rotation.  The average rms of the wavelength solution for the blue and red side of the DEIMOS chips were 0.25 and 0.27 km s$^{-1}$ respectively.  For the FORS2 spectra, the low signal to noise of the individual frames necessitated that
we perform cross correlation radial velocity calculations on the combined
spectra, rather than on each individual image.  As such heliocentric velocity 
corrections were tailored to the individual exposures and applied 
prior to combining the spectra, due to the long temporal 
baseline (roughly four months) of the FORS2 observations.  Once shifted and combined, the spectra were ready 
for radial velocity computation with the aid of template stars and a Fourier 
cross correlation routine (\textsc{fxcor}).  For the FORS2 data, a total of 23 template radial 
velocity stars observed with the same instrument setup were used with the 
cross correlation routine.  In the case of the DEIMOS data, the stellar spectra were cross correlated against a single synthetic template around the CaT region.  Both computations provided in-line error estimates, with the velocity errors ranging from $1.0 \leq \delta V_{hel} \leq 8.0$ km s$^{-1}$ with a mean of $\langle\delta V_{hel}\rangle = \pm 2.3$ km s$^{-1}$ for the DEIMOS stars, and a range of $3.0 \leq \delta V_{hel} \leq 10.0$ km s$^{-1}$ with a mean of $\langle\delta V_{hel}\rangle = \pm 5.0$ km s$^{-1}$ for the FORS2 stars.  For the FORS2 reductions systematic velocity 
errors due to a star's position in the slit were removed by centroiding the stars relative to the slit centre.  The fact that this procedure was done on combined spectra resulted in small absolute corrections, as the $\surd n$ statistics meant that the individual slit errors were minimized in the combination and correction steps.  The typical shift for the slit error on an individual exposure is approximately 6-9 km s$^{-1}$ for the FORS2 stars, and we note that this shift produces negligible uncertainties in the equivalent width error.  The final average absolute corrections to the slit centering errors on the combined spectra were on the order of $\leq 1.5$ km s$^{-1}$ for the FORS2 stars.


\subsection{Membership Determination}
\begin{figure}
\begin{center}
\ifpdf
\includegraphics[width=0.98\textwidth]{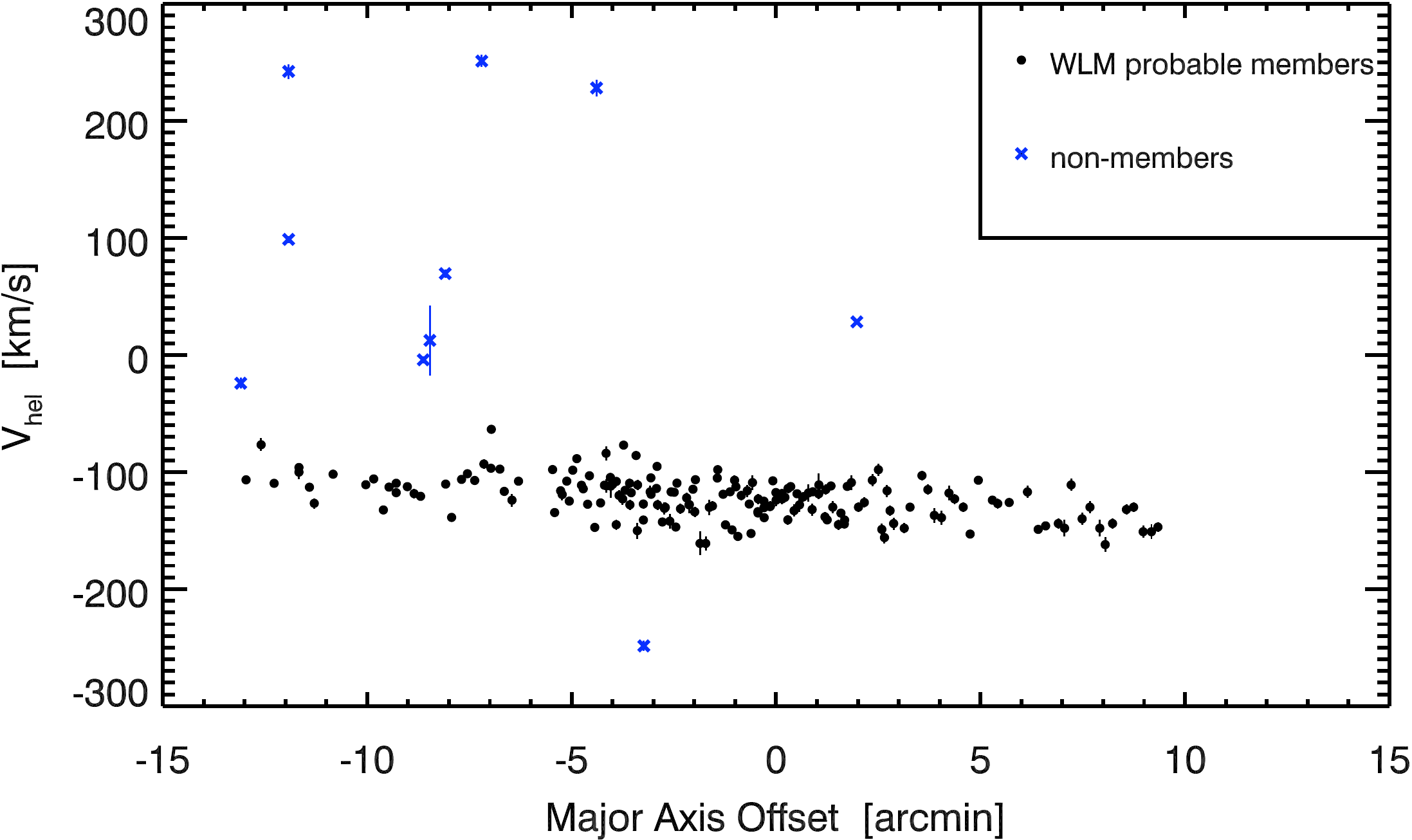}
\else
\includegraphics[angle=270,width=0.98\textwidth]{decvhelt.eps}
\fi
\caption{Heliocentric velocity as a function of major axis position for the 180 member RGB stars in WLM, as well as the non-member stars.  Extremely large absolute velocities are typically stars with very low signal to noise which rendered clean sky subtraction and radial velocity derivation difficult.  The mean random uncertainty on velocity for the member stars is 4 km s$^{-1}$ (smaller than the symbols in most cases). Due to the large distance off the plane of the galaxy ($b = -74^{\circ}$), few Milky Way stars contaminate our spectroscopic catalogue.}
\label{fig:vdf}
\end{center}
\end{figure}

\begin{figure}
\begin{center}
\ifpdf
\includegraphics[width=0.98\textwidth]{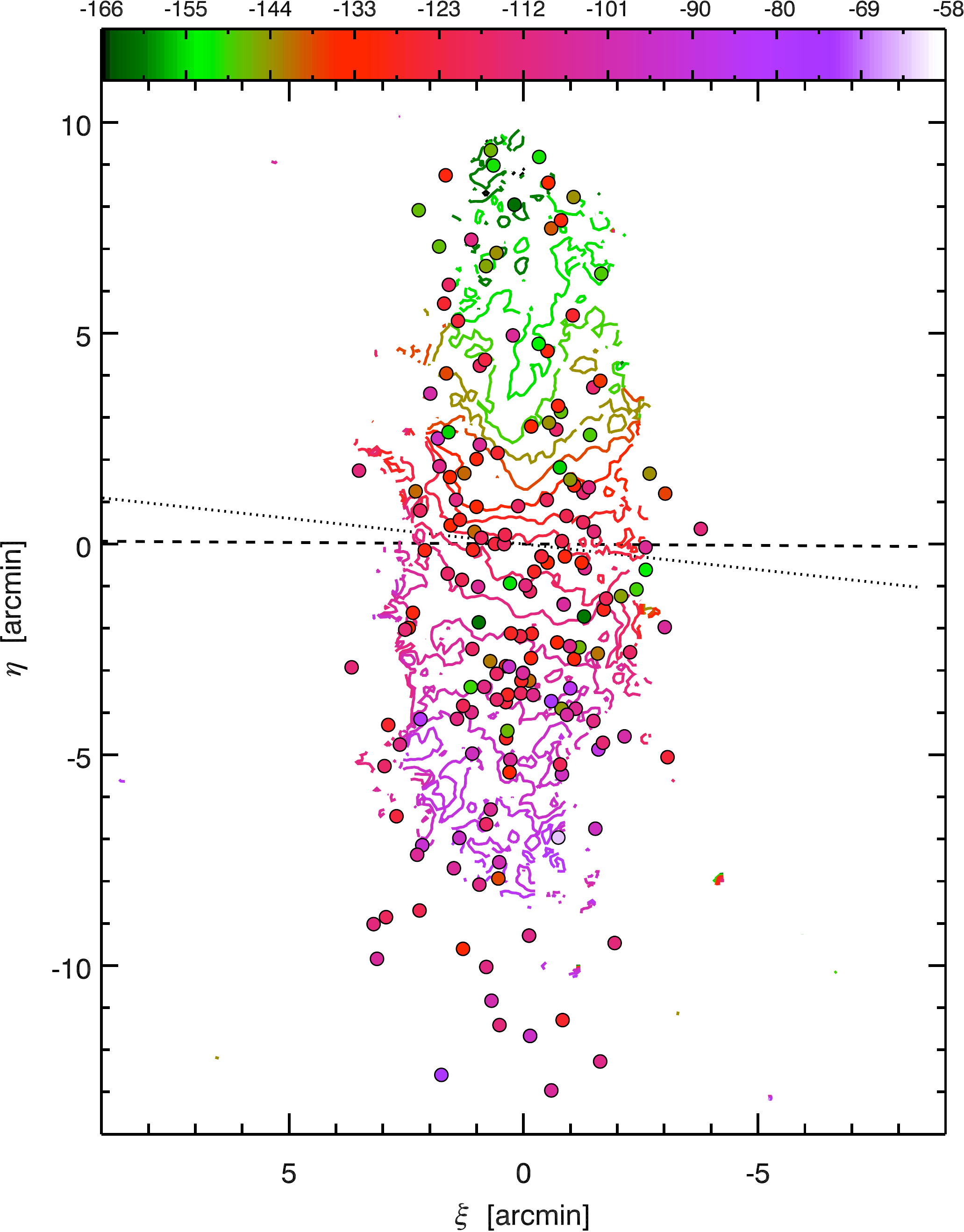}
\else
\includegraphics[width=0.93\textwidth]{gs2d.eps}
\fi
\caption{WLM spectroscopic sample in tangent plane coordinates on the sky.  HI velocity data from \cite{Kepley07} is shown as contours, and the stellar velocities of RGB members are colour coded to the same velocity scale. Dashed line shows the photometric minor axis, and the dotted line the kinematic angle of bisection determined as discussed in text.}
\label{fig:gs2d}
\end{center}
\end{figure}

Due to its distance off the plane of the Milky Way ($l,b = 75.85,-73.63$; \citealt{Gallouet75}), the contaminant fraction is expected to be very low in the direction of WLM.  Nevertheless, the DEIMOS sample of WLM covers a much larger field of view than the FORS2 sample presented in Paper I, which means the chance of foreground stars in the outer regions is increased.  Additionally, as we are possibly sampling member stars at larger radii, they stand to be at larger velocities with respect to the central regions.  Removing contaminants in a galaxy that is known to be rotating is not trivial, and in this case we culled the sample with an iterative maximum likelihood method.  Rather than a global $3-$sigma clipping, we adopted a spatially binned clipping routine which was much more robust, and by binning along the major axis avoided artificially inflating the $\sigma_{v}$ estimates.  This procedure was based on the joint maximum likelihood method of \cite{Walker06} (see also \citealt{GG79, Hargreaves94}),  which we extended here to spatial bins of 25 stars each, so as to accurately throw out contaminants when the rotational profile enhances the spread of velocities at different positions in the galaxy.  The algorithm simultaneously determines the dispersion ($\sigma_{v}$) and average velocity ($\langle u \rangle$) in a given bin by maximizing the probability:
\begin{equation}
\ln(p_{b}) = -\frac{1}{2}\sum_{i=1}^{N}\ln(\sigma_{i}^{2} + \sigma_{v}^{2}) - \frac{1}{2}\sum_{i=1}^{N}\frac{(v_{i}-\langle u \rangle)^{2}}{(\sigma_{i}^{2}+\sigma_{v}^{2})} - \frac{N}{2}\ln(2\pi).
\label{eqn:velml}
\end{equation}
 With this we culled the sample to 180 total member stars, with only one of the previous FORS2 stars removed.  Figure \ref{fig:vdf} shows the distribution of probable member and contaminant stars in our catalogue.  In Figure \ref{fig:gs2d} we plot the 2-D projected map of WLM with the HI gas contours and stellar sources determined to be members in our sample, both colour coded by velocity.  

\subsection{\textit{Spectroscopic Metallicities}}
As in Paper I, [Fe/H] derivations were done through the empirically calibrated Calcium II triplet method \citep{ADC91} whereby the summed equivalent width of these lines correlates with [Fe/H] on a scale calibrated to galactic globular clusters \citep{Cole04}.  We here use the recently updated calibration given in \cite{Starkenburg10}.  The details of the equivalent width measurements and placement onto the metallicity scale will be discussed in a forthcoming partner paper (Leaman et al., 2011b) which will deal with the chemical analysis and interpretation of WLM.  Of our 180 radial velocity members we have derived [Fe/H] measurements for 126 of the stars which had high enough signal to noise ($\gtrsim 10 \rm{\AA^{-1}}$).  The uncertainties were estimated in the fashion described in Paper I, with the average internal error on metallicity being $\Delta_{\rm{[Fe/H]}} = \pm 0.26$ dex.\\

\subsubsection{Repeat Observations}
While estimates of our errors exist for both [Fe/H] and velocity measurements, more information about how precise these measurements are can be gleaned from observations of stars that are in common between the FORS2 and DEIMOS dataset.  We ended up with 6 stars that overlapped between the sample for which to check velocities - and only 2 stars for which repeated [Fe/H] measurements can be made.  The values for these repeat measurements are shown in Table 2 .  The mean offset ($\langle X \rangle = \frac{1}{N}\Sigma (x_{i}-x_{j})$) and scaled median absolute deviation of the offsets ($S = 1.48\times med( {\mid}x_{i} - med(x_{j}){\mid} )$) are $X_{[Fe/H]} = 0.24;  S_{[Fe/H]} = 0.02$ dex, and $X_{Vel} = -2.5;  S_{Vel} = 13.9 $ km s$^{-1}$ respectively.  Given the velocity and metallicity uncertainties discussed in the preceding section, these repeat observations are consistent within the errors.

\subsection{Age Derivations}
At the distance of WLM, deriving ages for individual stars is not possible with precision greater than $\sim 50\%$, and in reality is simply providing a coarse correction to the assumption that metallicity and colour are perfectly correlated. The reader is referred to Paper I for the discussion of the age derivations in detail, but relevant updates are briefly outlined here.  The well calibrated homogeneous photometry from \cite{Alan05} is crucial in deriving accurate ages, but this method still suffers from difficulties (c.f., \citealt{Tolstoy03b}) - namely the possibility of internal differential reddening, and poorly constrained [$\alpha$/Fe] values for the RGB stars as discussed in Paper I.  Given the extended SFH of WLM \citep{Mateo98, Dolphin00} and the gas rich 
nature of dIrr galaxies, we expect a significant range in red giant ages 
in the dataset.  

With our current spectroscopic sample, [Fe/H] values are now found $\leq -2.5$ dex, which necessitated using the Yale-Yonsei evolutionary tracks \citep{Demarque04}, which sample much lower metallicities.  In Paper I we provided age estimates based on both [$\alpha$/Fe]$ = +0.3$ and solar.  Here we use a compilation of high dispersion spectroscopy of various stellar populations in WLM to constrain the mean [$\alpha$/Fe] as a function of metallicity.   \cite{Colucci10} observed the lone globular cluster (GC) of WLM, WLM-1 to have [Ca/Fe] = 0.25 and [Fe/H] = -1.71, which we adopt as the [$\alpha$/Fe] values for all our targets with similar or lower metallicities.  Of course GCs may undergo a separate formation and chemical evolution than field stars, however in the MW, LMC and Fornax the metal poor clusters show similar [$\alpha$/Fe] as the field stars \citep{Pritzl05, Hill97, Letarte06}, and Ca is not affected by GC mixing processes (e.g., \citealt{Gratton04}). Stars more metal rich than this are assigned an [$\alpha$/Fe] based on a third order polynomial, that is fit to the WLM-1 GC value, as well as the B, and A supergiant studies of \cite{Venn03} and \cite{Bresolin06}.  The position of each star on the colour magnitude diagram was interpolated among a grid of Yale-Yonsei isochrones at appropriate values of [Fe/H] and [$\alpha$/Fe] for the star.  

The random errors on ages for individual stars are $ \pm 50\%$.   Systematic errors are present in these estimates due to the variable $\alpha$-enhancement, extinction, and AGB contamination.  We have estimated the magnitude of these systematic uncertainties in our ages using artificial star tests, which will be described in more detail in the forthcoming paper (Leaman et al., 2011b).  In the relevant figures in this work we have overlaid both the individual random error in a star's age, as well as the combined systematic uncertainties.

\section{Results}
In this work we have added DEIMOS spectroscopic observations of 103 member stars to supplement the 77 member stars from Paper I.  We have determined the radial velocity 
values for these 180 \emph{individual} member RGB stars in WLM, of which 126 also have [Fe/H] and ages derived.    These new objects allow us to sample out to a projected radius of $3.4$ kpc ($\sim 3.5$ disk scale lengths) in WLM, assuming the adopted distance of 985 kpc.  This spatially extended sample can be used
to examine the structure, kinematics, and chemical evolution history of 
this galaxy for the first time in conjunction with derived age estimates.   
We now present an updated discussion of the structural and kinematic parameters.

\subsection{Structural Parameters}
\begin{figure}
\begin{center}
\ifpdf
\plottwo{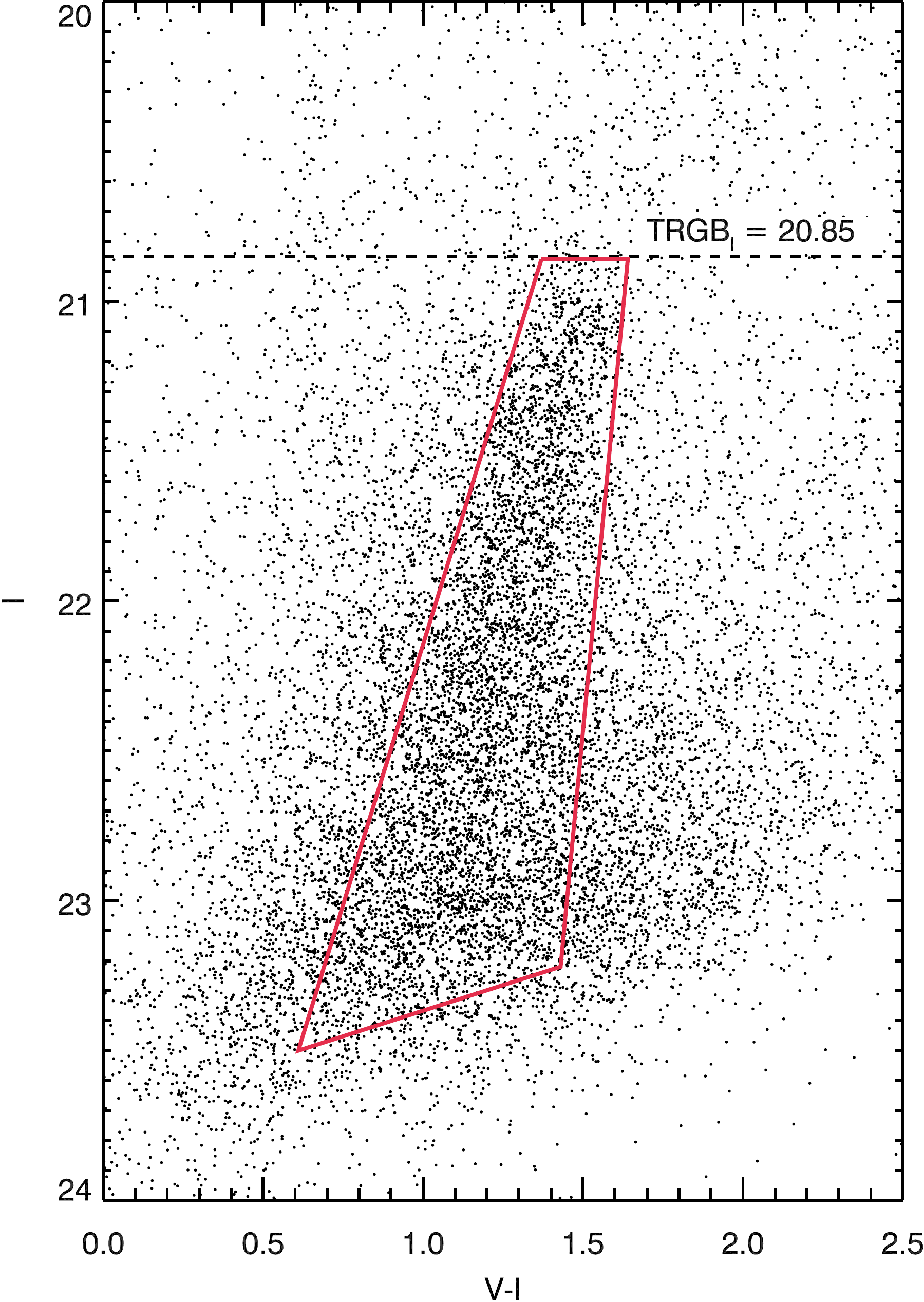}{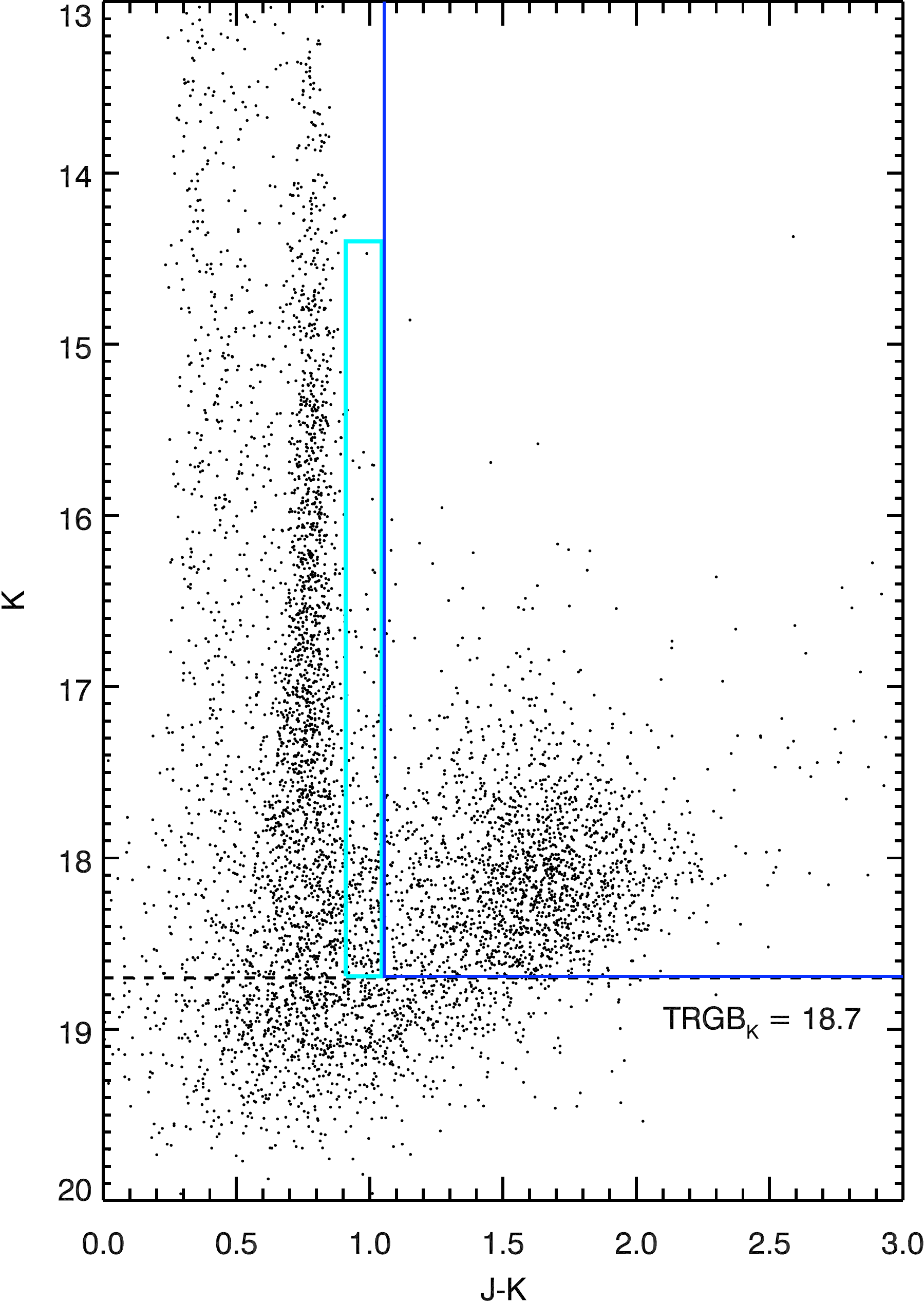}
\else
\plottwo{vicutn.eps}{jhkcutn.eps}
\fi
\caption{INT WFC V,I photometry, with strict RGB cut, presented with JHK UKIRT/WFCAM photometry of WLM \citep{Tatton10}.  The metallicity dependent cut between C-rich AGB stars (\emph{blue box}) and the O-rich AGB stars for WLM (\emph{light blue box}) is taken from \cite{Tatton10}.}
\label{fig:photcut}
\end{center}
\end{figure}

\begin{figure}
\begin{center}
\ifpdf
\includegraphics[width=0.98\textwidth]{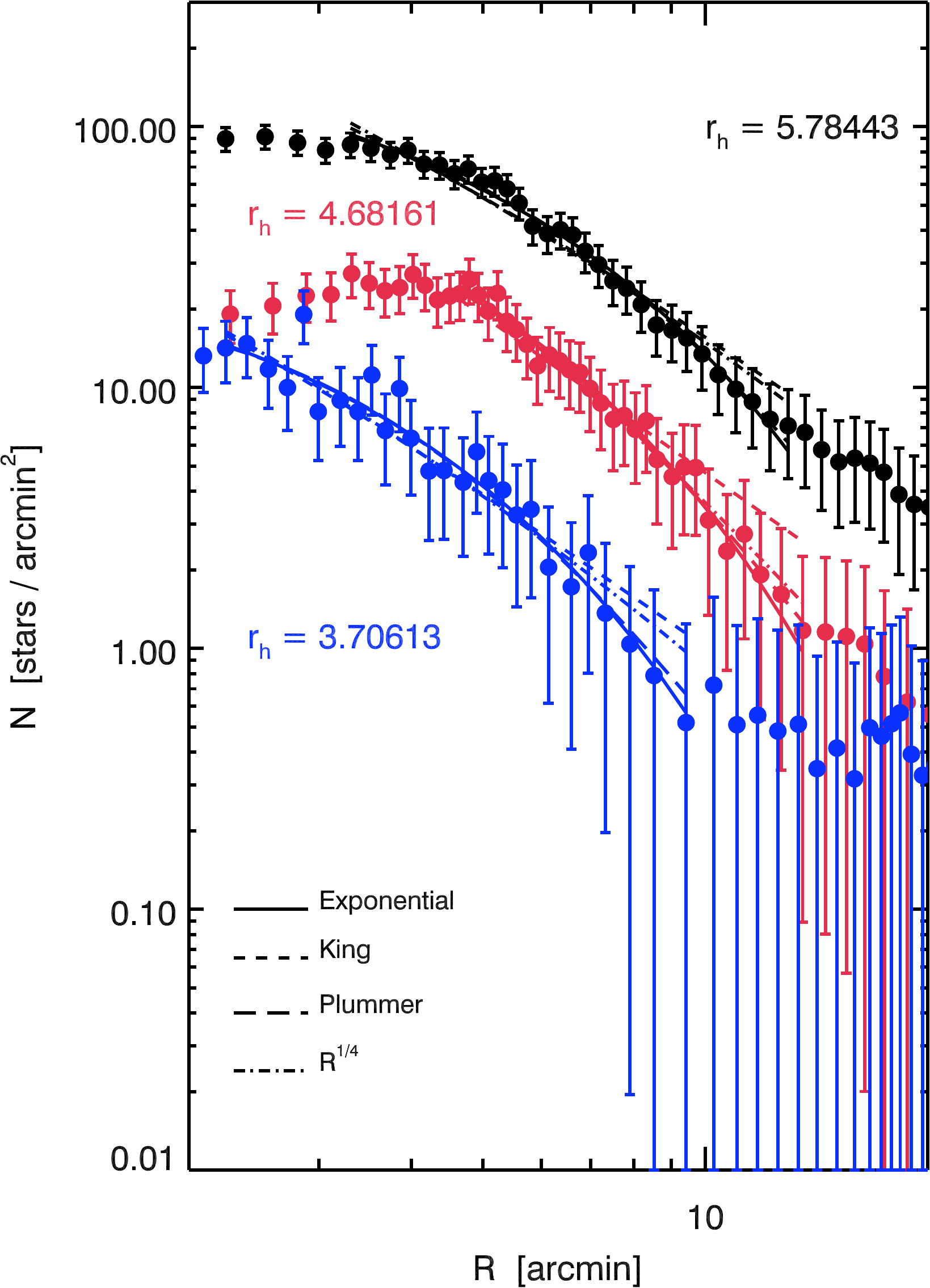}
\else
\includegraphics[width=0.95\textwidth]{stardense.eps}
\fi
\caption{Geometrical radial surface density profiles for the full V,I (black), strict RGB (red), and C-star (blue) stellar populations.}
\label{fig:stardense}
\end{center}
\end{figure}

To derive the structural parameters of WLM we used the photometric V,I, and JHK catalogues, as well as the HI data of \cite{Kepley07}.  The colour magnitude diagrams are shown in Figure \ref{fig:photcut}, including the INT V, I photometry as well as the infrared UKIRT/WFCAM photometry of WLM.  We have isolated several population tracers, as shown by the coloured areas in the figures.  For the V, I photometry  a strict RGB locus is shown by the red box.  The tip of the RGB location was taken from the analysis of \cite{Alan05} which used the same photometric catalogue as our present work.  This selection represents the least contaminated red giant branch population in WLM.  Outside of this lie \emph{primarily} excessively young or metal poor RGB and luminous AGB stars.  In the JHK diagram the oxygen rich M-stars (\emph{light blue}) and the young ($\sim 1$ Gyr) carbon rich C-stars are identified, as those AGB stars above the TRGB and to the appropriate side of the metallicity dependent colour cut found for WLM in the work of \cite{Tatton10}.  The stellar density profiles for the full V,I (\emph{black}), strict RGB (\emph{red}), and C-star (\emph{blue}) populations are shown in Figure \ref{fig:stardense}, constructed by binning stars in geometrical radii which have been referred to an average ellipticity of $e = 0.55$ at a position angle of $179$ degrees.  The inner regions of the two INT populations suffer from crowding, and so the subsequent fits of the surface density profiles exclude these regions.  Four profiles fits have been applied; Plummer \citep{Plummer11}, King \citep{King62}, de Vaucouleurs and exponential profiles.  For the rest of this paper we adopt the best fitting exponential surface brightness profiles, $\Sigma(R) = \Sigma_{0}e^{-R/R_{d}}$, and convert the exponential scale length, $R_{d} = 3.45^{'}$ (for the full V,I sample), into the 2D projected half light radius, $r_{h} = 5.78^{'}$, and where necessary the 3D deprojected half light radius, $r_{h3} = 7.68^{'}$, using the relations of \cite{JWolf10}.  The background level in each population was simultaneously fit as a constant, and was found to be consistent with selecting out by eye isolated regions of our field of view to estimate the background level.

\begin{figure}
\begin{center}
\ifpdf
\includegraphics[width=0.80\textwidth]{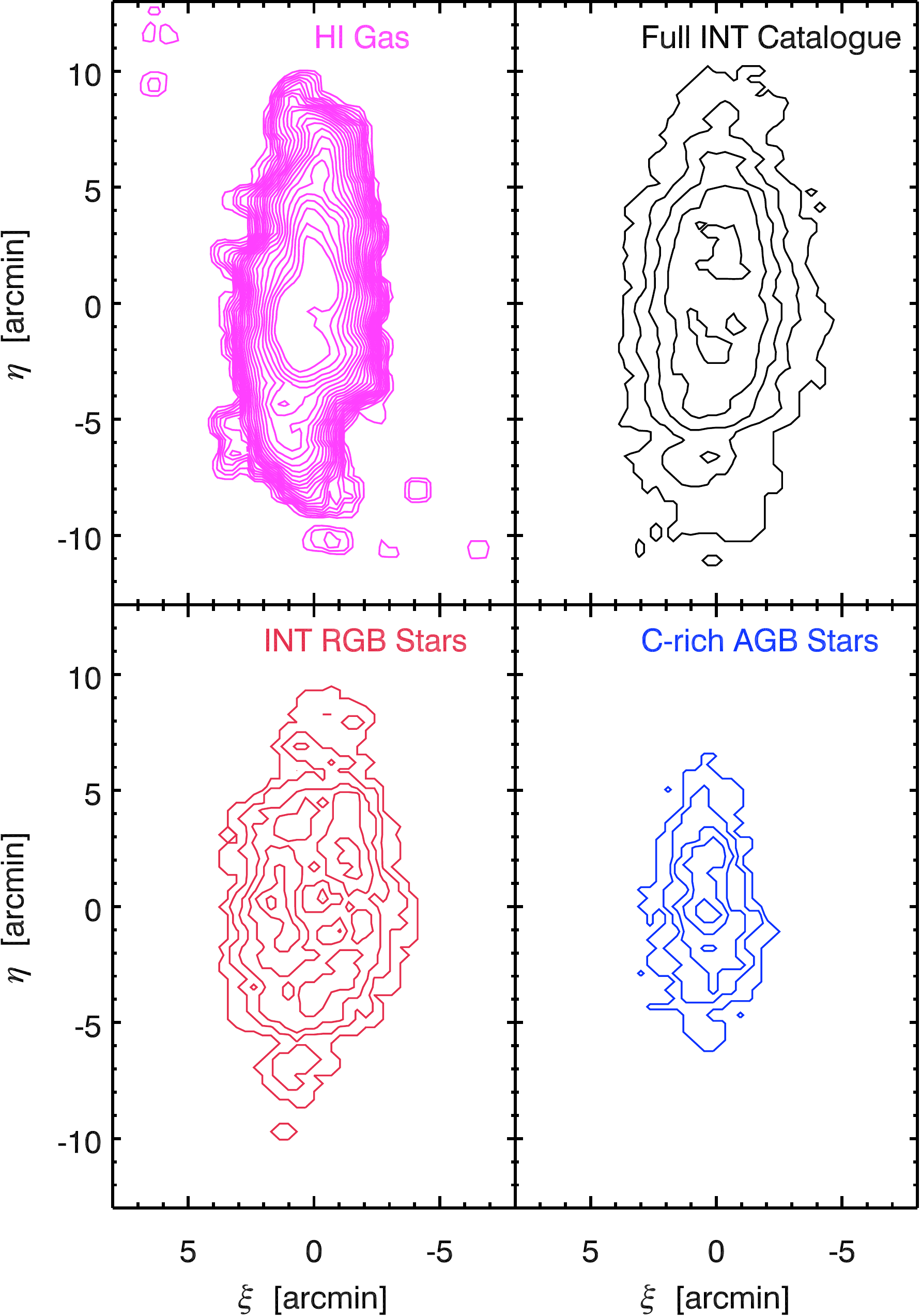}
\else
\includegraphics[width=0.80\textwidth]{4isopleth2.eps}
\fi
\caption{Density contours for the stellar and gaseous populations of WLM in the tangent plane.  Contour levels start at $2\sigma$ above the background and each subsequent contour level is $n$ times higher than the last, where $n = 1.2$.}
\label{fig:4cont}
\end{center}
\end{figure}

The 2D projected spatial distributions of these three populations and the HI surface density are shown in Figure \ref{fig:4cont}.  The mean and standard deviation of the background level was determined from regions at the outer limits of our field of view.  Contour levels start at $2\sigma$ above the background and each subsequent contour level is $n = 1.2$ times higher than the last. The HI radial extent is not significantly larger than the stellar component as reported in \cite{Kepley07} and the recent single dish observations of WLM by \cite{Hunter11}.  The latter work also noted that the density profile of the HI shows a smooth drop off to the detection limit, with no sharp truncation.  The strict RGB population shows a slightly more spherical distribution than the HI, but otherwise the general shape of the evolved stellar populations are consistent with the HI population.  As expected the young C-star population occupies the central regions of the galaxy, where the most recent star formation has occurred \citep{Dolphin00}.  These centrally concentrated younger populations are similar to what has been found in other Local Group dwarf galaxies (e.g., \citealt{Battaglia06, Hidalgo09, DeBoer11}).  However we note that with characteristic ages of $\sim 1$ Gyr, these stars would still have experienced $\sim 3$ rotation periods.    Ellipticity and photometric position angle, ($\rm{PA_{phot}}$), are calculated as a function of radius by fitting ellipses using the task \texttt{MPFITELLIPSE} \citep{MPFIT09}, on the contours at various radii for each population.  

\begin{figure}
\begin{center}
\ifpdf
\includegraphics[width=0.98\textwidth]{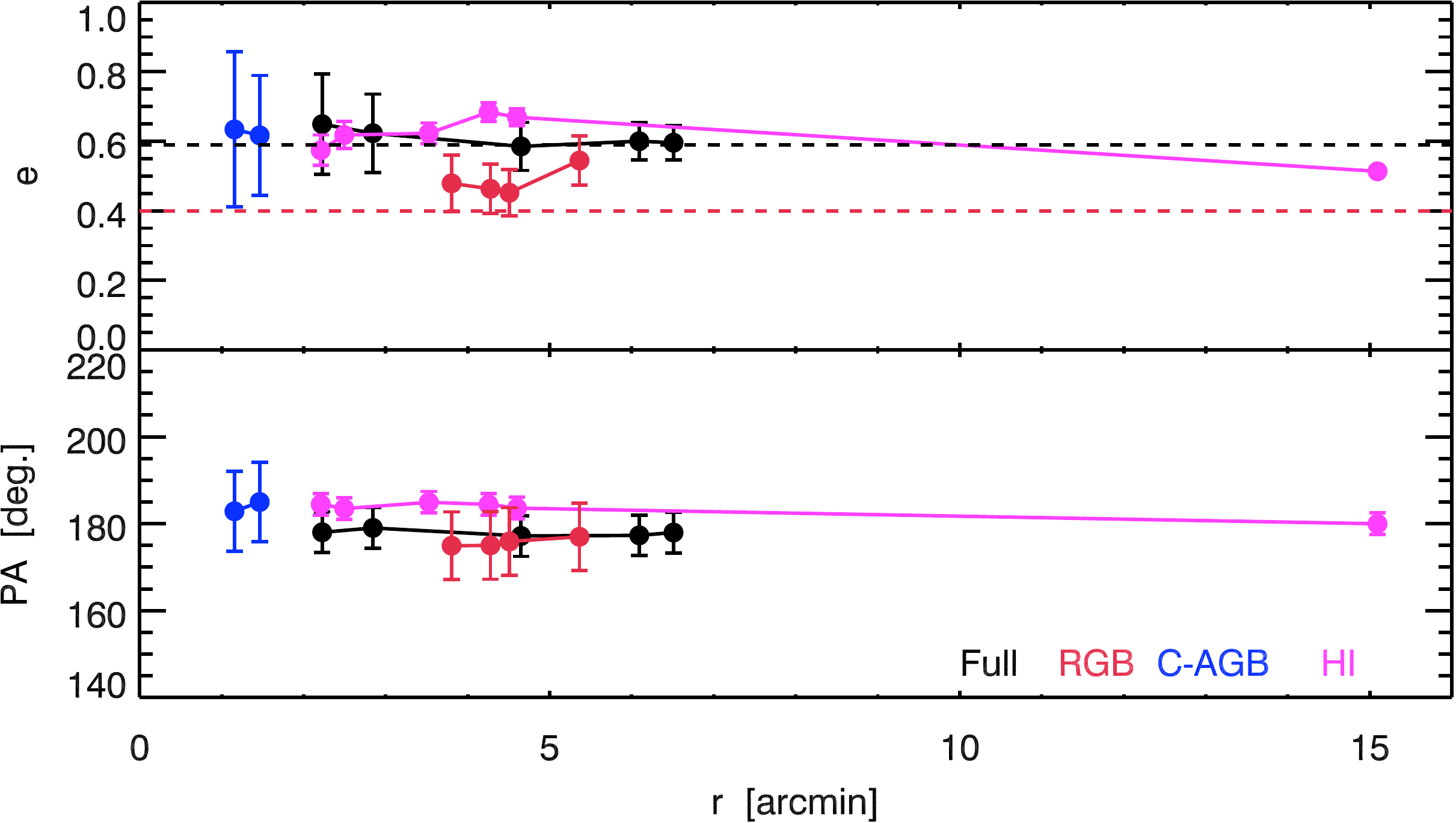}
\else
\includegraphics[angle=270,width=0.98\textwidth]{paerell2.eps}
\fi
\caption{Derived ellipticity and photometric position angle as a function of major axis radius for the various populations. Black is the full INT catalogue of stars, red the strict RGB population, blue the C-rich AGB stars, and magenta the HI gas.  Dashed lines show the measured disk ellipticity from \cite{Ables77} in black, and extended RGB halo ellipticity of \cite{Minniti97} (\textit{red}).}
\label{fig:psipaerell}
\end{center}
\end{figure}

The resulting position angles and ellipticities are shown as a function of radius in Figure \ref{fig:psipaerell}.  There is excellent agreement between the HI and stellar structural properties with only the evolved RGB population showing a slightly rounder ellipticity.  This is consistent with literature values of past photometric studies of \citep{Ables77,Minniti97}.  The close agreement in position angle and ellipticity of the stellar populations in WLM is in contrast to some lower angular momentum dSphs such as Phoenix and Fornax \citep{MD99,Battaglia06}.  WLM clearly is flattened more than the nearby dSphs in the Local Group \citep{Mateo98} - which may be due to the strong rotation as we shall see in the next section\\
\subsection{Internal Kinematics}
With nearly equal rotation and dispersion in the stellar component, care must be taken in interpreting WLM's velocity data.  In rotating systems, classical mass estimates rely on the HI gas, which has negligible random motion, making for relatively simple analyses equating the enclosed mass to some circular velocity.  The underlying mass profile can also be ascertained through velocity dispersion profile based methods - such as spherical Jeans modelling (e.g., \citealt{Battaglia08}).  In each of these cases, only one type of dominant kinematics is present in the tracer population - rotation or dispersion.  The challenge in our case is to convert WLM's composite dynamical configuration into a singular rotating or dispersion dominated dataset.  Without these corrections a number of systematic biases arise when interpreting rotational and pressure supported velocity distributions - such as encountered when analyzing an object in the configuration of WLM (c.f. \citealt{Lokas10}).  Detailed dynamical modelling will be presented in a forthcoming paper (Leaman \& Battaglia, in prep.).  However the analysis of rotation and velocity dispersion profiles used in the remainder of this work requires accurate removal of the biases that mixed rotation/dispersion systems show, which we discuss now.

\subsubsection{Stellar Velocity Field}
\begin{figure}
\begin{center}
\ifpdf
\includegraphics[width=0.98\textwidth]{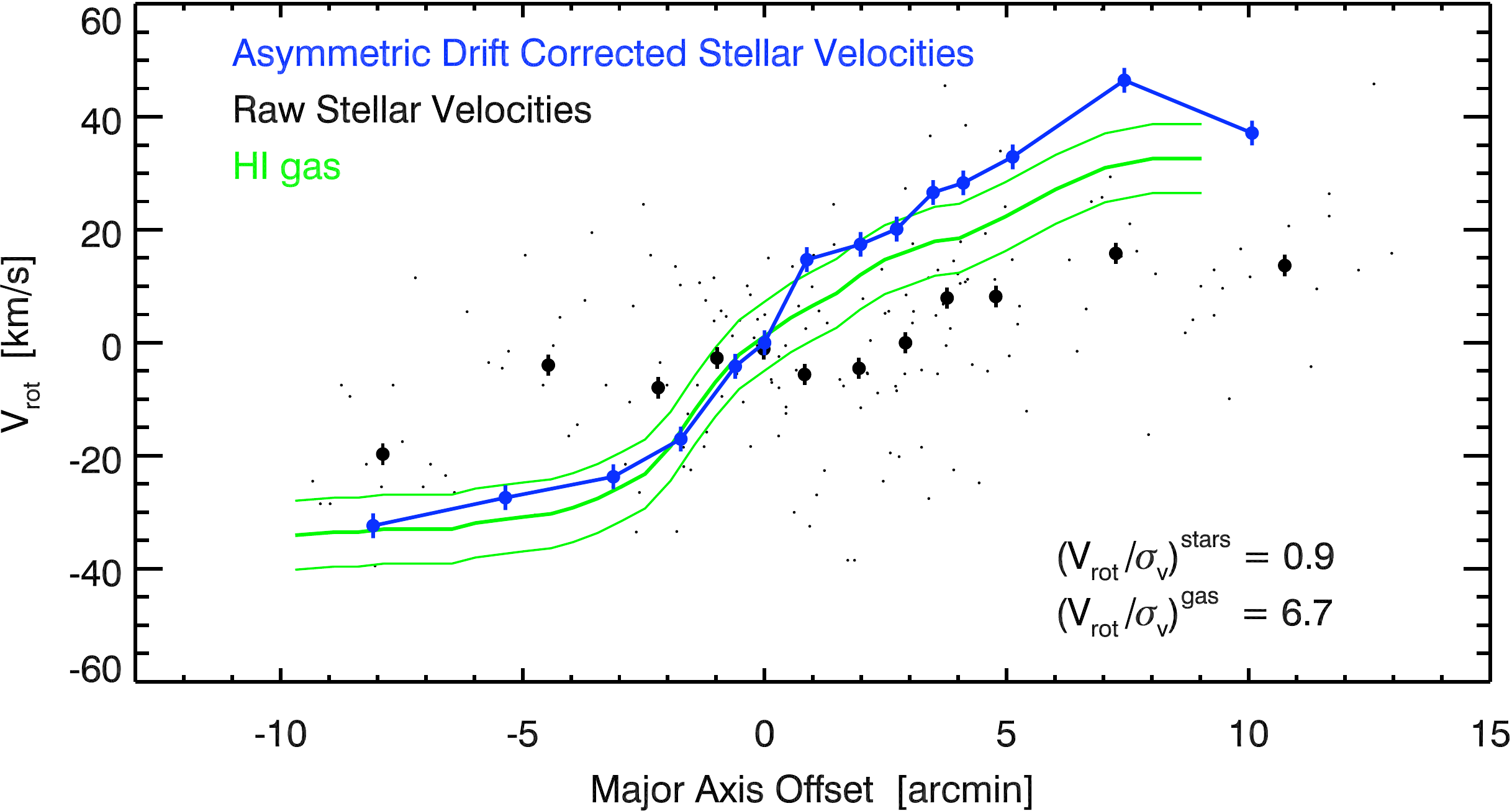}
\else
\includegraphics[angle=270,width=0.98\textwidth]{sgrotn.eps}
\fi
\caption{WLM stellar and HI rotational velocities ($V_{hel} - V_{sys}$) as a function of major axis offset. Small black dots represent the individual stellar data points, and the large circles the binned values with error on the mean shown.  Results of the asymmetric drift correction to the stellar rotation is shown in blue.  Also noted are the relative rotation over pressure support for the raw stellar and gaseous kinematics.}
\label{fig:gsrot}
\end{center}
\end{figure}

\begin{figure}
\begin{center}
\ifpdf
\includegraphics[width=0.90\textwidth]{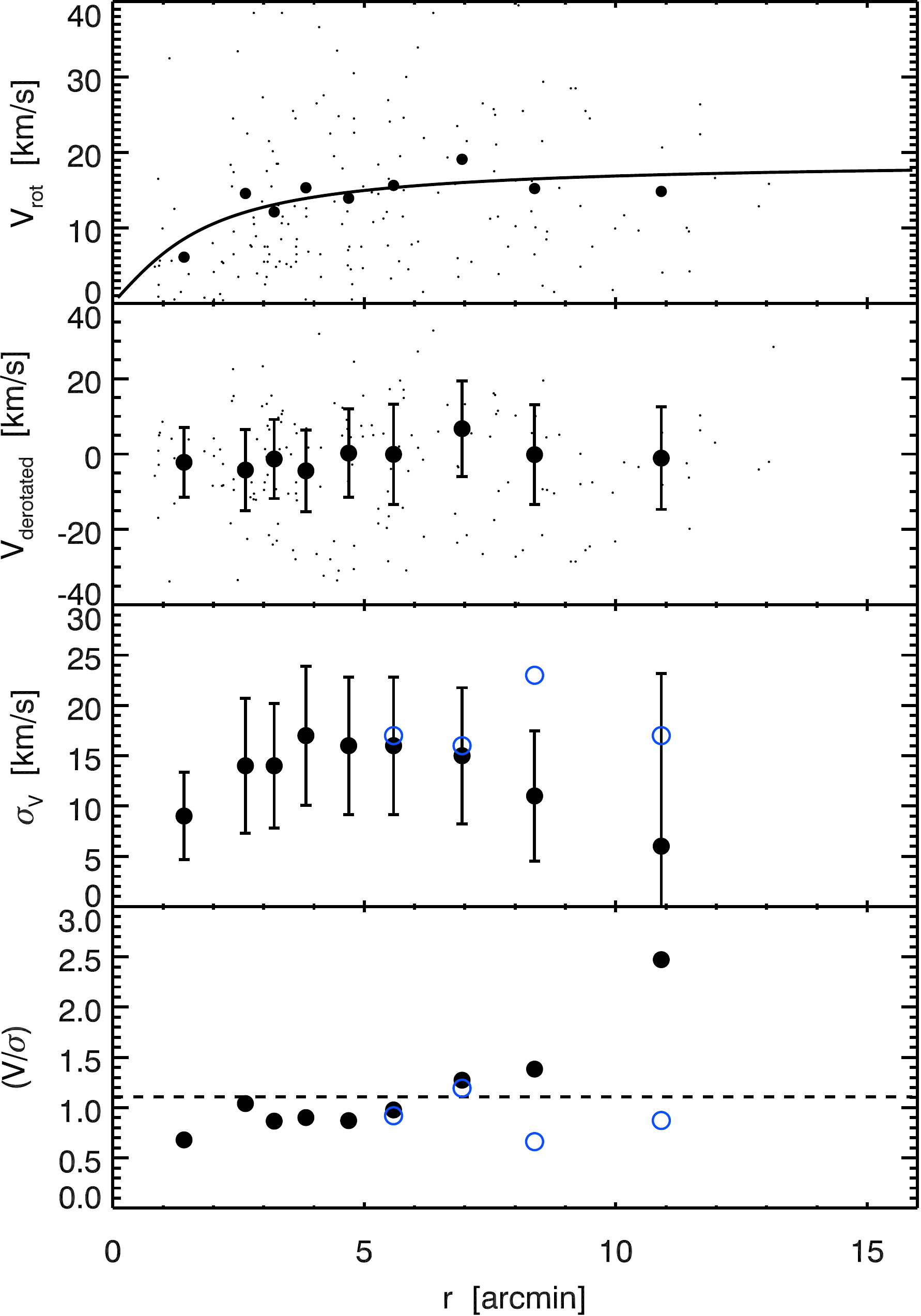}
\else
\includegraphics[width=0.80\textwidth]{wlmprofnetest.eps}
\fi
\caption{Raw stellar rotational velocity (\emph{top}), along with the optimal derotation achieved through a single fit through the major axis velocity profile of WLM. Third panel shows velocity dispersion profile before (\emph{blue circles}) and after rotation is subtracted - with the artificial enhancement of $\sigma_{v}$ at large radii removed in the latter case.  Bottom panel shows ratio of rotational to pressure support, with dashed line showing the expected value for an oblate galaxy flattened by isotropic rotation to the ellipticity of WLM.}
\label{fig:velprof}
\end{center}
\end{figure}

As discussed in Paper I, the stellar velocities are markedly different in WLM from most of the dSph galaxies.  The stars in WLM of all ages/metallicities appear to be rotating, in contrast to the purely random stellar motions in Local Group dSphs.  We begin by looking at the velocity structure of the stellar and gaseous populations of WLM.  As shown in Figure \ref{fig:gsrot}, WLM's stellar velocities, while showing rotation greater than seen in the dSphs of the Local Group, rotate with half the velocity of the HI gas, and lag behind the HI by $\sim 15$ km s$^{-1}$.  Clearly the stellar rotation is decoupled from that of the HI, with the suppressed rotation velocity of the stars due to the effect of asymmetric drift as we will demonstrate later.  In Figure \ref{fig:velprof} we show the projection of the rotation velocity of the stars as a function of geometric radii (again referred to an ellipticity of $e = 0.55$).  The binned data points are fit with an isothermal sphere rotation curve of the form $V(r)^{2} = 4{\pi}a_{0}a_{1}^{2}( 1  - a_{1}/r)({\rm atan}(r/a_{1}))$ \citep{Kepley07}, which slowly rises to a value of $ \sim 17$ km s$^{-1}$ at the last measured point.

As later we wish to derive a line of sight velocity dispersion profile, it is important that we accurately characterize the velocity field of WLM.   We therefore explored different models to characterize the rotation of the stars in WLM.  A few methods that were tested were: fitting the stellar velocity field simply along a strip of the major axis, describing the rotation field by iterating through fits at differing position angles, or treating the stars with the classical ``spider diagram'' velocity field:
\begin{equation}
v(x,y) = v_{sys} + v_{rot}(R)cos(\theta)sin(i) + v_{rad}(R)sin(\theta)sin(i),
\label{eqn:spider}
\end{equation}
where $\theta$ is the azimuthal angle and $i$ the inclination of the galaxy.

The most consistent determination, as judged by the smallest deviation at all radii in the rotation-subtracted velocities, was a simple linear fit along the velocity pattern of the major axis.  The rotation subtracted velocities are shown in the second panel of Figure \ref{fig:velprof}, with the mean values of the stars staying close to the systemic velocity at all radii.  The velocity of the stars off the major axis shows little deviation from the model, suggesting that the stars are rotating cylindrically - however due to the uncertainties in the inclination and disk thickness of WLM we can not examine this in detail.

\subsubsection{Velocity Dispersion Profile}
  Deriving a line of sight velocity dispersion ($\sigma_{v}$) profile for these stars requires removal of the rotation signature, so that the $\sigma_{v}$ profile is not artificially enhanced at large radii.   As we showed above, WLM exhibits significant rotation, so correcting for this magnitude of rotation in a galaxy with this structure is not trivial.   Using the best fit velocity field model described in the previous paragraph, the rotation was removed and we could proceed to construct a velocity dispersion profile that was unbiased at large radii.  The dispersion was calculated applying the likelihood method of Equation 1, using the rotation subtracted velocities.  Errors on the dispersion estimates were taken to be the value at the $1-\sigma$ confidence interval of the likelihood parameter space.  The resulting dispersion profile is shown in the third panel of Figure \ref{fig:velprof} as solid black dots.  The open blue circles in that figure show the resulting $\sigma_{v}$ profile if we were to compute the dispersion profile on the raw velocities that \textit{hadn't} been rotation subtracted.  The artificial enhancement of the dispersion values at large radii is clearly evident due to the strong streaming motion of the stellar velocity field.  The radial velocity dispersion profile for WLM does not appear to be flat, unlike what is seen in many dSphs \citep{Walker09}, however interestingly the dE NGC 147 which is of similar mass, luminosity, and shape as WLM shows a drop off in $\sigma_{v}$ at larger radii  as well \citep{Geha10}.  This and other nuances of the dispersion profile will be discussed in depth in a forthcoming paper on the detailed mass profile of WLM (Leaman \& Battaglia, in prep.).  In the bottom panel of Figure \ref{fig:velprof} we show the ratio of rotational to dispersion support in WLM as a function of radius, $V/\sigma$.  The dashed line represents the expected value of $V/\sigma$ for an oblate isotropic rotator at the eccentricity of WLM.  The flattening of WLM then is most likely due to rotation, rather than anisotropy in the outer regions.  The value of $V/\sigma \sim 1$ within one effective radius is much less than the HI, which shows $V/\sigma \sim 7$.  Nearly equal pressure and rotation support for the stellar component is also much lower than what simulations typically use for progenitor dIrrs (i.e., \citealt{Mayer01a,Kazantzidis11}).  The measured $V/\sigma$ may reflect some minor interactions between WLM and other dwarfs in the Local Group, however as discussed in the next section, the alignment of the kinematic and photometric axes of WLM rules out significant interactions.  Therefore if this value of $V/\sigma = 1$ for the isolated WLM dIrr is representative of typical dIrr progenitors, it would have implications for the efficiencies (and thus timescales) of dwarf galaxy tidal transformation scenarios in such simulations. 
\\
\subsection{Photometric and Kinematic Axes}
\begin{figure}
\begin{center}
\ifpdf
\includegraphics[width=0.98\textwidth]{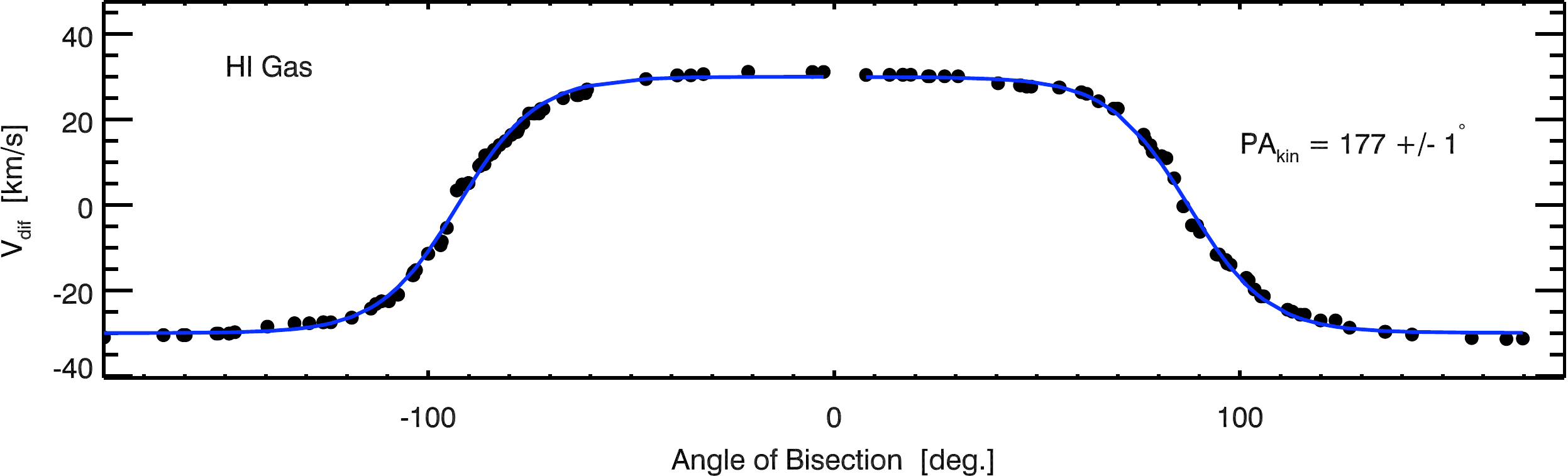}
\includegraphics[width=0.98\textwidth]{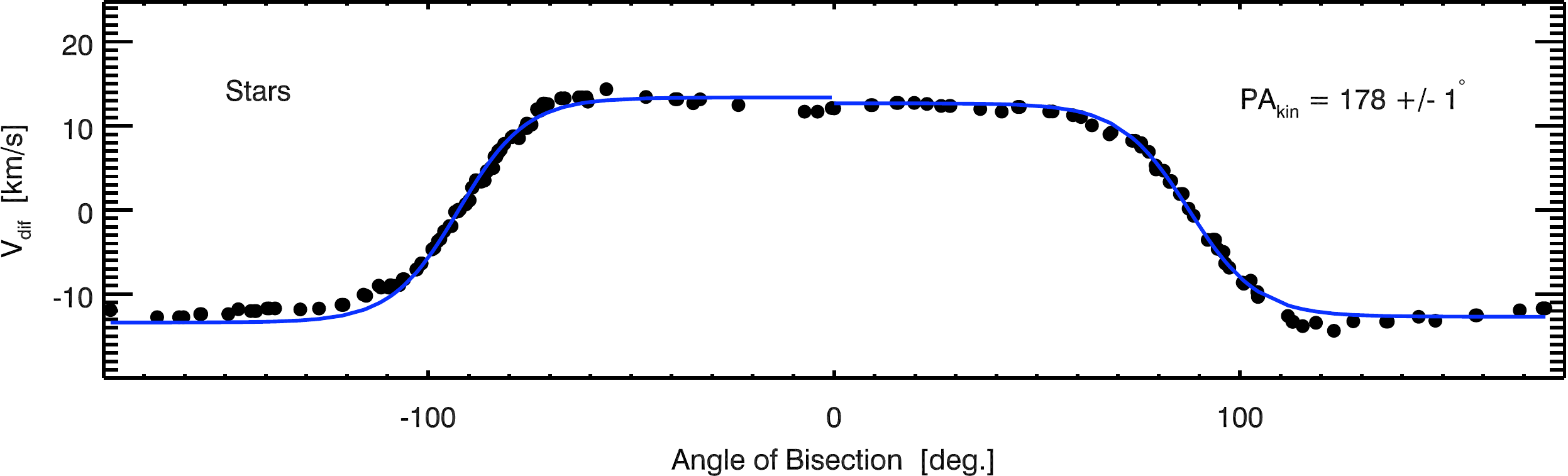}
\else
\includegraphics[angle=270,width=0.98\textwidth]{kinpag2.eps}
\includegraphics[angle=270,width=0.98\textwidth]{kinpas2.eps}
\fi
\caption{Kinematic position angle diagnostic plots for the HI (\emph{top}) and stellar data (\emph{bottom}).  The velocity difference (see text) between the stellar sample on either half of a bisecting axis is plotted as a function of the bisection angle.  Here the bisection angle of zero degrees corresponds to the minor axis of WLM.  Overlaid is a function of the form $\rm{a_{0}tanh}(a_{1} (\theta-a_{2}))$.}
\label{fig:pakin}
\end{center}
\end{figure}

Information on the orientation and potential of WLM comes from comparing the position angle of the kinematic ($\rm{PA_{kin}}$) and photometric ($\rm{PA_{phot}}$) data.  The kinematic position angle defines the major axis of rotation of a galaxy, and in cases where recent mergers have occurred or the observer is viewing a triaxial system, may not always lie along the same axis as the photometric position angle.  To compute the $\rm{PA_{kin}}$ we proceed as in e.g., \cite{Walker06}.  For every star, or interpolated HI velocity point in the tangent ($\xi,\eta$) plane, an angle of bisection is calculated based on the data point's position off the above determined photometric minor axis.  The galaxy is split along this bisection line and velocity difference is computed by calculating the average velocities above and below the line and subtracting them.  The bisection line is then rotated by one degree for the HI or to the next star and the procedure is repeated, building up the curves shown in Figure \ref{fig:pakin}.  A hyperbolic tangent function was fit to the data and the position angle determined based on the bisection angle where the velocity difference is minimized. With a position angle of zero northwards in the tangent plane on the sky, the kinematic position angle is defined as 180 degrees + or - the angle of bisection where the minimum velocity difference occurs.  The uncertainty on the angle is determined by calculating the difference from the bisection line that would be produced if the velocity difference was changed by the average uncertainty in a stars velocity. We find consistent values for the kinematic position angle of the stellar and gaseous components of WLM, with the latter values in excellent agreement with the HI studies of \cite{Jackson04} and \cite{Kepley07}, who used other methodology.  

  With the photometric and kinematic position angles calculated, it is instructive to look at their relative alignment, typically designated as the parameter $\Psi \equiv  |PA_{kin} - PA_{phot}|$.  The quantity of $\Psi$ is sometimes used as a signature of mergers, where the structural and dynamical misalignment would be high, or to give insight into non-circular gaseous motions due to infall or outflow \citep{Fraternali04}.  For galaxies in the ATLAS 3D study $\sim 90$\% of galaxies were found to have alignments $\leq 15^{\circ}$ \citep{Krajnovic11}.  For the stellar component of WLM we derive $\Psi = 1 \pm 3^{\circ}$, and $\Psi = 2 \pm 2^{\circ}$ for the HI.  The fact that the alignment of the kinematic and photometric position angles is close, also constrains the viewing angle and shape.  As discussed recently by \cite{Lokas10}, when rotation is significant in a triaxial system viewed perpendicular to the longest dimension, rotation other than the major axis is usually evident, as the likelihood of the rotation being aligned with either axis is small in a potential that permits more irregular orbits.  In contrast an axisymmetric system will show no rotation along the minor or intermediary position angles, thus the kinematic and photometric position angles will show strong alignment like we observe in WLM.  Additional constraints against WLM being triaxial come from its position on the $V/\sigma$ vs. $e$ anisotropy diagram.  As we shall see in the next section WLM is flattened mostly by rotation rather than anisotropy, consistent with the interpretation that we are viewing a highly inclined oblate spheroid.

\subsection{How Thick is WLM?}
\begin{figure}
\begin{center}
\ifpdf
\includegraphics[width=0.98\textwidth]{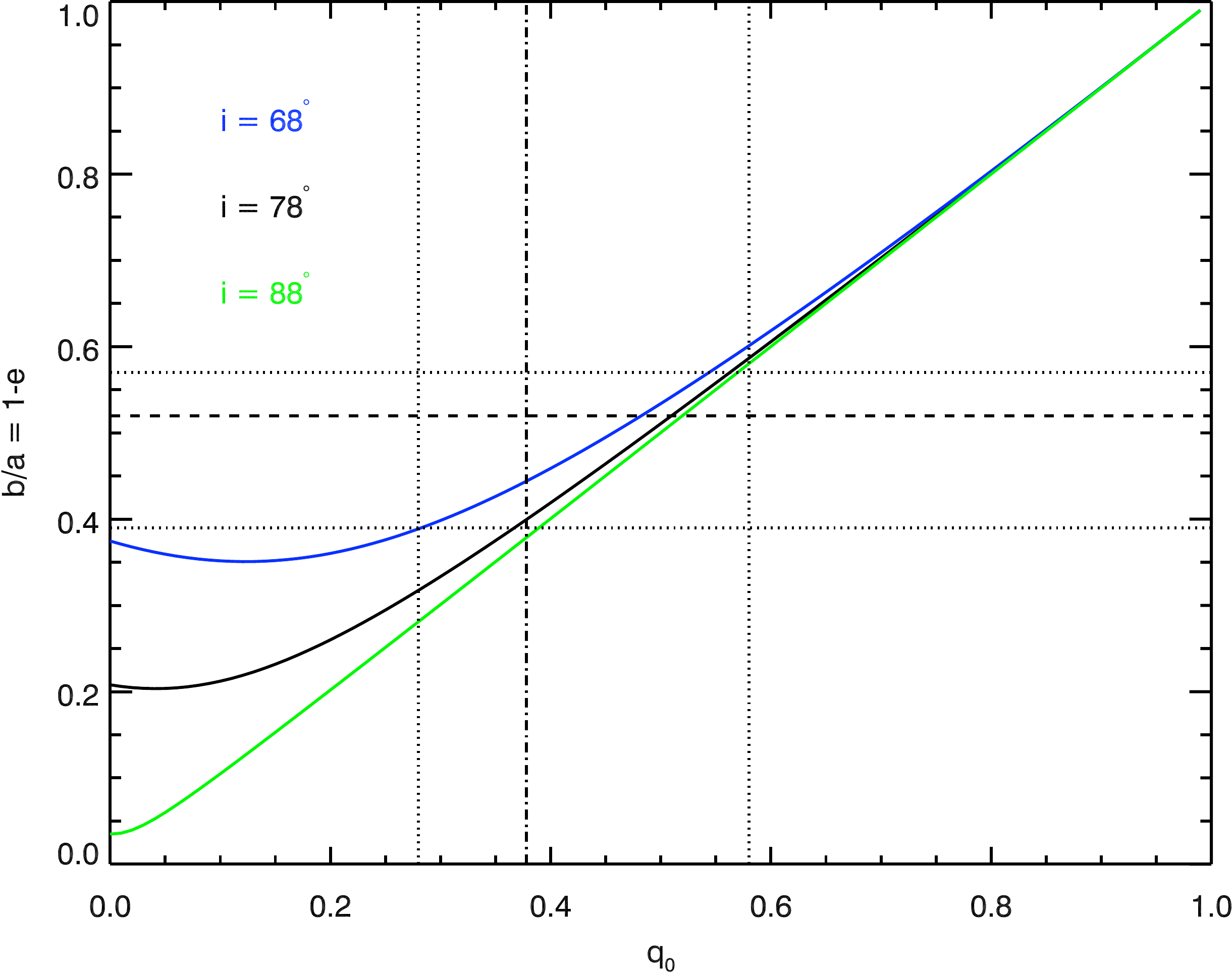}
\else
\includegraphics[angle=270,width=0.98\textwidth]{qvba.eps}
\fi
\caption{Range of expected intrinsic thickness for an observed projected axial ratio.  Shown are curves for the mean, upper and lower inclinations from the HI tilted ring analysis of \cite{Kepley07}.  Horizontal dotted lines represent the range of axial ratios from the isopleth analysis presented in Figure \ref{fig:psipaerell}, with the horizontal dashed lines the mean values for the stellar populations.  Vertical dotted lines indicate the corresponding range of $q_{0}$ values given the inclination limits from the HI analysis.  Vertical dashdot line indicates the intrinsic axial ratio assuming the measured axial ratio from the minor and major axis fits of Figure \ref{fig:tdcomp}.}
\label{fig:qvba}
\end{center}
\end{figure}

From Figure \ref{fig:psipaerell} the ellipticity and position angle of the gas and stars allow us to characterize the projected distribution of those populations.  To untangle the intrinsic shape of WLM is more complicated.  If we are observing an oblate spheroid that is rotationally flattened, it will have an intrinsic thickness,  $q_{0}$ (for this orientation, $q = c/a$, with $c$ being the short axis and $a$ being the long axis) - which is difficult to disentangle without a constraint on the viewing inclination, $i$.  The relation between the intrinsic and projected axial ratio is shown in Figure \ref{fig:qvba} for the range of inclinations determined in the tilted ring analysis of the HI data in \cite{Kepley07}.  We note that because their inclinations were derived assuming a thin disk in the tilted ring analysis, they therefore provide lower limits on the possible inclination.  Thus given the high inclination for WLM, the range of measured projected axial ratios $(b/a)$ for the stellar and gaseous components (horizontal lines), suggest that the intrinsic thickness ranges between $0.38 \leq q_{0} \leq 0.57$.  With the data presented here, we have evidence that WLM's vertical structure is quite extended in both its stellar and gaseous components.  This is confirmed using the techniques described in \cite{vdM02}, whereby WLM can be approximated as a flattened spheroid with axial ratio $q(R) = ( 2 + [\frac{2R_{D}}{R} (\frac{V_{rot}(R)}{\sigma(R)})^{2}])^{-1/2}$, assuming isotropic velocities for the stars (which are assumed to lie in a larger isothermal halo).  Using the derived exponential scale length from $\S4.1$, the intrinsic axial ratio ($q$) for WLM calculated using this method again lies between 0.38 and 0.56 at all radii - in good agreement with Figure \ref{fig:qvba}.

\begin{figure}
\begin{center}
\ifpdf
\includegraphics[width=0.98\textwidth]{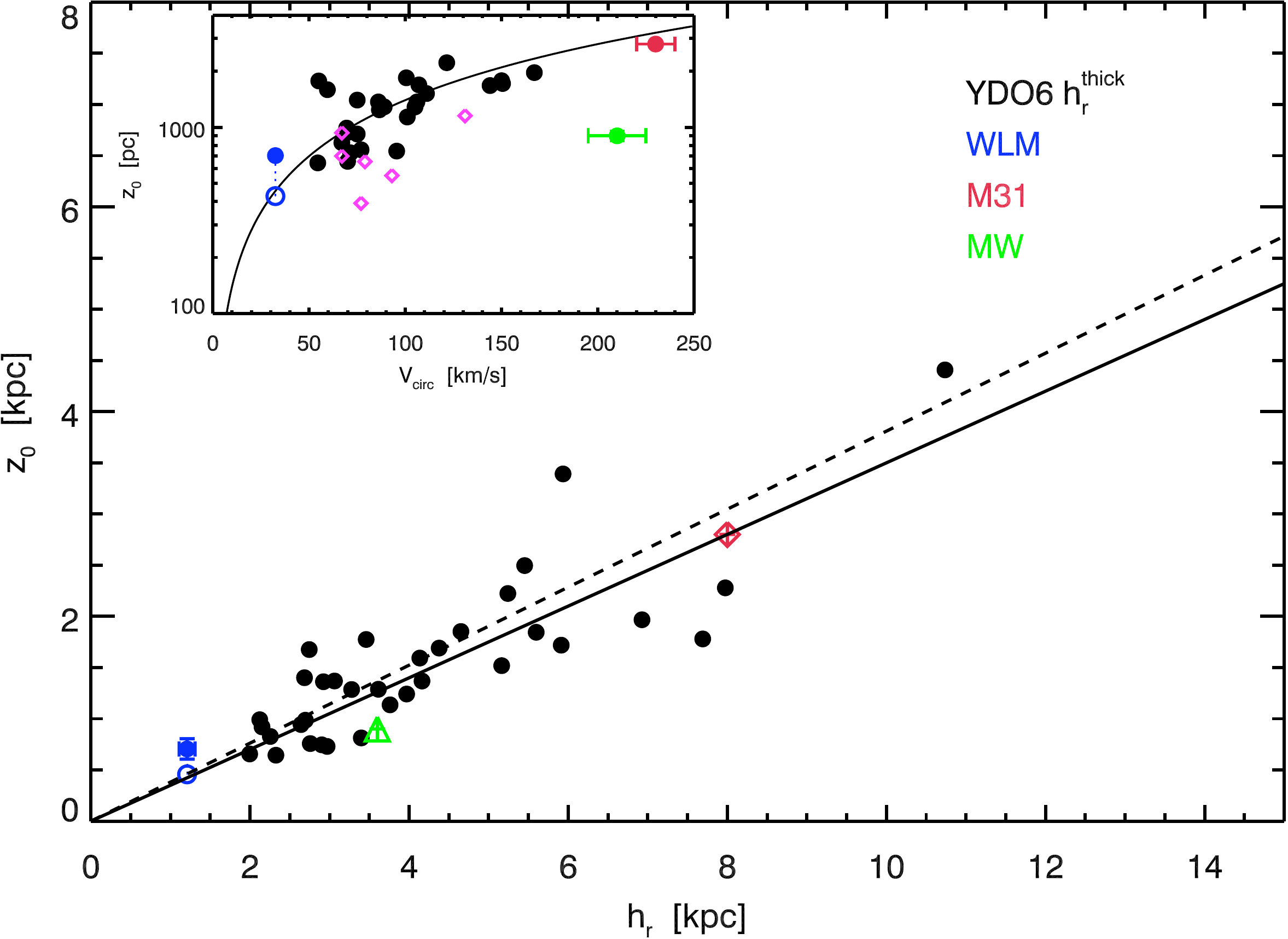}
\else
\includegraphics[angle=270,width=0.98\textwidth]{tdcomp.eps}
\fi
\caption{Comparison of major axis radial scale lengths versus vertical thick disk scale heights.  Black dots are data from the work of \cite{YD06}, and M31 and MW data taken from \cite{Collins11} and references therein.  Inset shows vertical scale heights in parsecs as a function of $V_{circ}$ for the same data plus 6 resolved stellar population studies from \cite{Seth05}.  Black line is a fit of the form $Z_{0,thick} = (1400 pc)(\frac{V_{c}}{100 {\rm km s}^{-1}})^{1.0} $ adopted by \cite{YD06} for their galaxy sample.  Open circle for WLM shows the effect of the inclination correction as discussed in the text.}
\label{fig:tdcomp}
\end{center}
\end{figure}

This value is consistent with the recent study of \cite{Roychowdhury10}, who showed that the average intrinsic \emph{gas disk} thickness of dwarf irregulars is a much higher $\langle q_{0} \rangle = 0.6$, instead of the 0.2 that is canonically assumed for stellar and gaseous disks in higher mass galaxies.  Our values for the stellar thickness are also in line with studies of distant dwarf galaxies that seem to show thicker stellar structure \citep{SJ10}.\\
\\
We might ask how does this thick configuration compare to other galaxies where direct measurements of the vertical height are possible?  While we are unable to determine whether the extended vertical structure in WLM is due to just a single component flattened spheroid, or multiple disks of varying thickness, we can compare the vertical extent to studies of edge on galaxies that exhibit thick disks.  In Figure \ref{fig:tdcomp}, we plot the radial and vertical scale heights for WLM along with a sample of edge on spiral galaxies from \cite{YD06}, and the two large spirals in the Local Group (c.f. \citealt{Collins11}).  The values for WLM were derived by fitting the \emph{major axis} stellar surface density with an exponential profile, and the minor axis with a profile of the form $\Sigma(z) \propto \rm{sech}^{2}(z/z_{0})$, for which we derive a vertical scale height of $z_{0} = 705 \pm 28$ pc. Assuming an inclination of $i = 79$ we show the corrected vertical scale height marked on the figure as the open circle.  The best fit lines are taken from \cite{YD06} and indicate that WLM has a similar relative vertical scale height as expected for its radial scale length.  Direct confirmation of a distinct thick disk in WLM would only be possible after identifying a velocity lag or age difference in a statistical sense, which is beyond the scope of our spectroscopic sample.  However the vertical structure of WLM is clearly extended to an amount consistent with other galaxies that show increased vertical scale heights.

\section{Mass Estimates}
Examination of the velocities shows that WLM is clearly in a different structural and dynamical state than dSphs, however does that mean their mass estimates also differ?  Detailed mass modelling can provide answers to both what the stellar velocities tell us about the past history of WLM, and about the underlying potential.  As we have both stellar and HI velocities there exists the opportunity for much more detailed modelling than is available from dSph data sets.  With this we can compare how various mass modelling techniques perform on a system where there is strong rotation and flattening.  Various dynamical models to describe the mass profile of WLM will be presented in a forthcoming paper but for now we can compute simple mass estimates in two ways.

\subsection{Rotationally Derived Mass}
We can use the inherent rotation of the stars to compute an estimate of the circular velocity of WLM, provided we remove the component of random motions that are suppressing the rotation curve.  As noted earlier, the stellar rotation curve differs in magnitude to the HI due to the effect of asymmetric drift.  To directly compare the rotation curves of the stars and gas, and allow for independent circular velocity estimates we must correct for this effect. To analyze the stellar kinematics in a purely rotational sense corrections due to asymmetric drift were applied.
\begin{equation}
v_{c}^{2} = v_{\phi}^{2} + \sigma^{2}(r)\left[\frac{\partial \rho}{\partial R} + \frac{\partial \ln\sigma_{R}^{2}}{\partial \ln R} + (1-\frac{\sigma_{\phi}^{2}}{\sigma_{R}^{2}}) + R\sigma_{Rz}^{2}\frac{\partial(\ln\rho\sigma_{Rz}^{2})}{\partial z}\right].
\label{eqn:ad}
\end{equation}
Here $v_{\phi}$ is the inclination corrected observed rotation velocity, $\rho$ is the stellar density, $\sigma$ the velocity dispersion in a given component, and $v_{c}$ is the drift corrected rotation velocity.  We follow the outline and assumptions of \cite{Hinz01}: namely that the last term can be neglected, $\frac{\sigma_{\phi}^{2}}{\sigma_{R}^{2}} = 0.5$, the density profile is of the form $\rho \propto \rm{exp}(-R/R_{d})$, and the slope of the \emph{radial} component of the velocity dispersion is flat in the outer regions of the disk.  The equation then reduces to $v_{c}^{2} = v_{\phi}^{2} + \sigma_{\phi}^{2}(2R/R_{d} - 1)$.  We note that the assumed anisotropy is chosen primarily because it reproduces flat rotation curves in large galaxies, however we have little in the way of constraints on it at this time.  We use a nominal inclination of $i = 79^{\circ}$ for the inclination corrected values of dispersion and rotation, allowing us to compute the corrected stellar circular velocity, $v_{c}$.  As noted earlier, the tilted ring analysis of \cite{Kepley07} assumed a thin disk, therefore their inclination which we adopt here should provide a conservative lower limit.  The results of this correction are seen in Figure \ref{fig:gsrot}.  The final inclination and asymmetric drift corrected curve shown by the blue binned points matches well with the HI gas - giving us confidence that our dispersion measurements are correct for the stars. 

\begin{figure}
\begin{center}
\ifpdf
\includegraphics[width=0.98\textwidth]{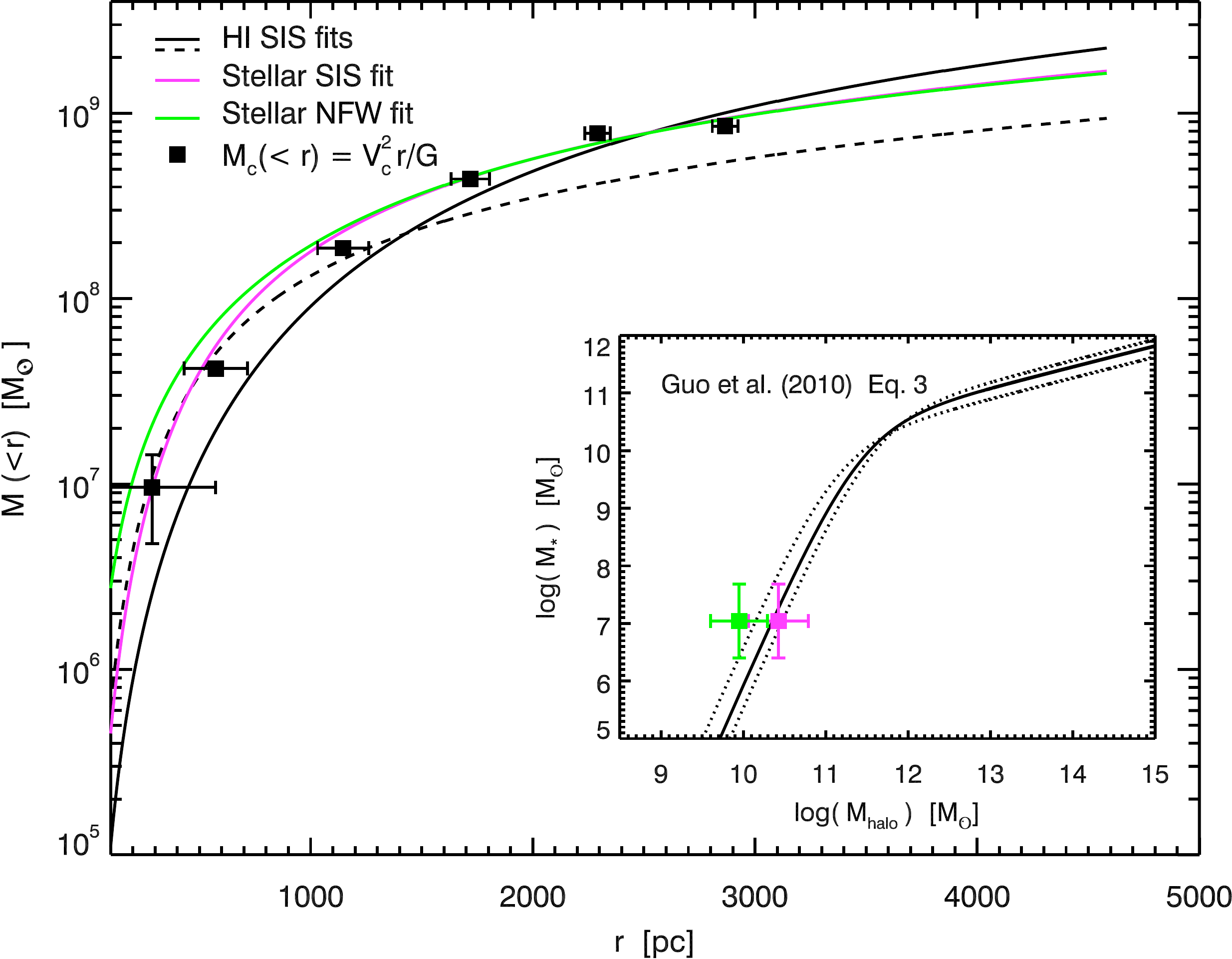}
\else
\includegraphics[angle=270,width=0.98\textwidth]{enclmassn3.eps}
\fi
\caption{Enclosed Mass as a function of geometrical radius.  Black squares are rotationally derived mass estimates using the asymmetric drift corrected stellar velocities at various radii.  Black lines show the HI rotation curve isothermal sphere estimates from the best fit models of \cite{Kepley07} to the receding and approaching sides of the galaxy.  Cyan curve is the best fitting isothermal sphere model fit to the six stellar data points (see text).  Green curve is NFW mass model fit to the same points.  Inset shows stellar mass as a function of the implied halo mass from both the SIS and NFW models, with comparison to the relation (\emph{black line}) from \cite{Guo10} and references therein.}
\label{fig:enclmass}
\end{center}
\end{figure}

With this corrected stellar rotation curve we have estimates of the circular velocity at various projected radii in WLM, with which we can compute the associated rotationally derived mass as $M(<r) = \frac{V_{c}^{2} r}{G}$.  In Figure \ref{fig:enclmass} we show these dynamical mass estimates as a function of radius as the black squares.  The black solid and dashed lines show the enclosed mass curves based on the HI rotation analysis from the study of \cite{Kepley07}, assuming an isothermal sphere model.  Agreement between the HI and stellar derived mass is roughly consistent when considering both the receding and approaching sides of the HI data.  Differences most likely result from the uncertainty in assuming a velocity anisotropy for the stellar circular velocity reconstructions, and possibly any non-circular motions not taken into account in the HI analysis.  A weighted best fit isothermal sphere model based on the six stellar bins is shown in magenta and an NFW mass model in green.  The rotationally derived mass at the half light radii ($1656 \pm 49$ pc) from the stellar data is $M_{half} = 4.3 \pm 0.3 \times 10^{8}$ M$_{\odot}$, and $M_{last} \simeq 8.5\times10^{8}$ M$_{\odot}$ at the last measured stellar bin ($2865$ pc).  Assuming the best fitting enclosed mass profile from the HI study of \cite{Kepley07}, one finds masses of $M_{half} \simeq 3\times10^{8}$ M$_{\odot}$, and $M_{last} \simeq 7\times10^{8}$ M$_{\odot}$ from the gas rotation at those same radii - in close agreement with the values from the stellar rotation.  The NFW and isothermal sphere models that are fit to the stellar data imply halo masses (taken to be the mass interior to a radius where the enclosed density equals 200 times the critical density) of $8.9\times10^{9}$ M$_{\odot}$ and $2.6\times10^{10}$ M$_{\odot}$ respectively.  We show these values plotted versus the known stellar mass of WLM ($1.1\times10^{7}$ M$_{\odot}$; \citealt{Jackson07a}) in the inset of Figure \ref{fig:enclmass}, along with the $M_{*} - M_{halo}$ relation of \cite{Guo10}.  The isothermal sphere model and NFW model are offset from each other, but both roughly consistent with the relation from \cite{Guo10}.  However the sensitivity of the derived halo mass to the rotation signature at small radii in WLM prevents us from discussing the implications of this in any further detail in this work.  With the above values and assuming a V band luminosity of $5.2\times10^{7}$ L$_{\odot}$ \citep{Mateo98}, we find a mass to light ratio of $M/L = 8$ and $29$ M$_{\odot}$ L$_{\odot}^{-1}$ at the half light radius, and 4.5 kpc (roughly the extent of the stellar and gaseous body of WLM) respectively. 

\subsection{Dispersion Based Mass}
An alternative estimate of the dynamical mass can be taken from simple analytic expressions relating the mass to the line of sight velocity dispersion at a characteristic radius, as in the recent works of \cite{Walker09}, \cite{JWolf10}, and \cite{Amorisco11}.  Independently, the studies identified a characteristic radius where velocity anisotropy effects are minimized.  With these formulae the enclosed mass is computed in accordance with simple virial estimates.  The \cite{JWolf10} relation requires a luminosity weighted velocity dispersion, which may introduce bias when applied to composite populations which have spatially different luminosity distributions (c.f., \citealt{Amorisco11}).  Therefore we use the relation of \cite{Walker09} who found $M(r_{h}) = \mu r_{h} \sigma^{2}$, where $\mu \equiv 580 \rm{M_{\odot}pc^{-1}km^{-2}s^{2}}$.  Being easily computed  this is used frequently to derive masses in the dSphs and UFDs of the Local Group, however there are several points of caution to be noted in this formula's application to WLM.  Any of these dispersion based mass estimators assume a system that is spherically symmetric, without rotation, and with a flat radial velocity dispersion profile - all conditions which are violated to some degree in WLM.  Nevertheless it is instructive for the time being to compare the dispersion based mass to other estimators.  We leave discussion of the biases that violation of these assumptions creates to our forthcoming paper on the dynamical modelling of WLM.  With these caveats we proceed and compute the average velocity dispersion within $r_{h}$ using the rotation subtracted velocity dispersion profile shown in Figure \ref{fig:velprof}.  With this the half mass from the \cite{Walker09} relation is $M_{half} = 2.1\pm0.3\times10^{8}$ M$_{\odot}$.  In Table  3 we summarize the mass estimates for WLM for ease of comparison.

\begin{figure}
\begin{center}
\ifpdf
\includegraphics[width=0.95\textwidth]{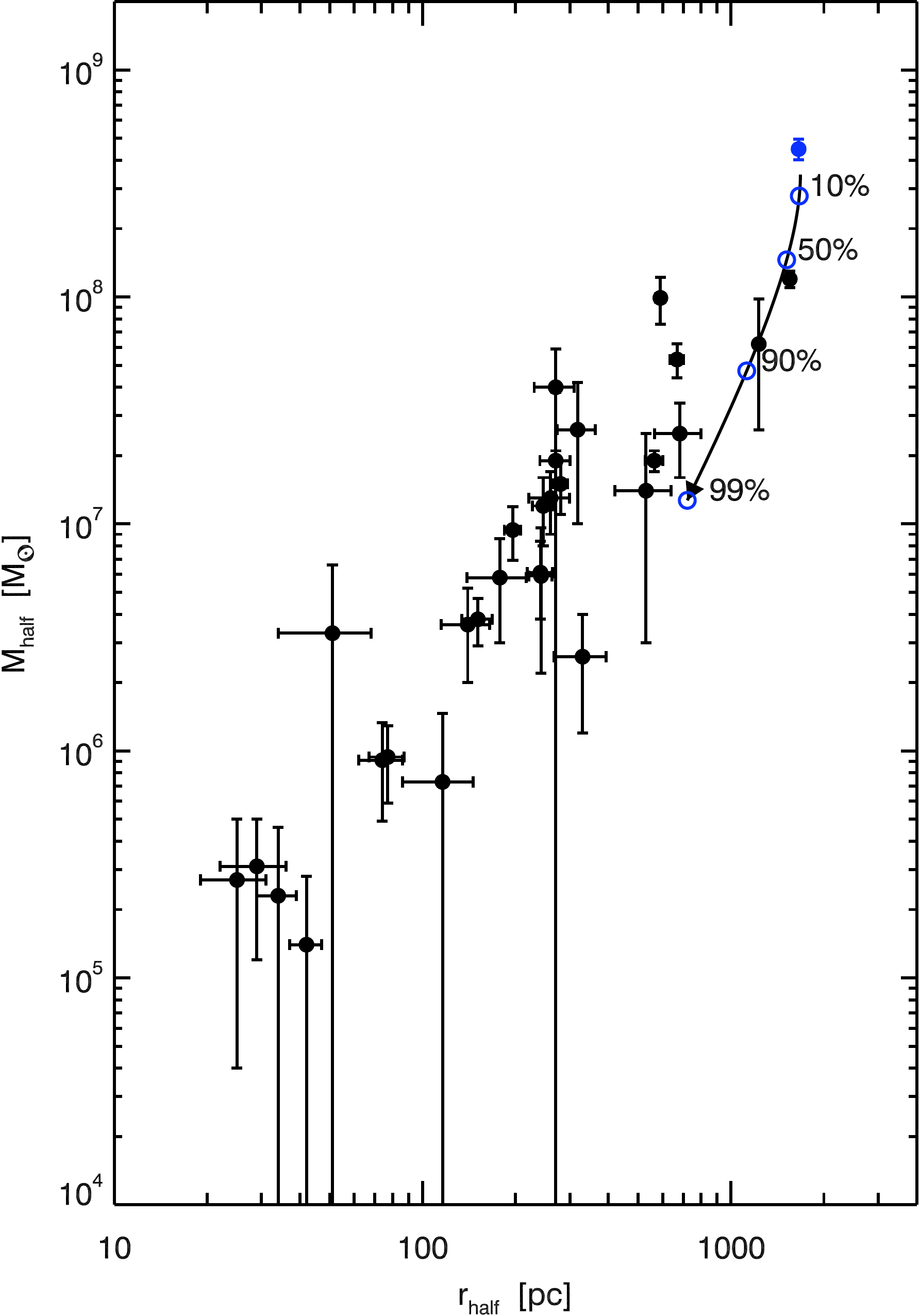}
\else
\includegraphics[width=0.85\textwidth]{mhalfrhalf.eps}
\fi
\caption{Half mass as a function of half light radius for Local Group dSphs, and WLM (blue).  Overlaid are the expected tidal evolution tracks of Penarrubia et al. 2008 for various total stripping values of the initial luminous mass.  WLM lies on the isolated side of the relation which is consistent with its implied separation and orbital history (see $\S 1.1$).}
\label{fig:mhalfrhalf}
\end{center}
\end{figure}

With the enclosed mass at the half light radius it is now instructive to compare WLM to the half mass estimates of Local Group dSphs.  In Figure \ref{fig:mhalfrhalf} we show that WLM lies slightly on the more massive side of  the sequence of $M_{half}$ vs. $r_{h}$ shown in \cite{Walker09}.  The other dSphs of the Local Group that lie off the relation in this direction are the isolated Cetus and Tucana dSphs.  Tidal evolution tracks of \cite{Penarrubia08} are overlaid, showing the expected evolution in mass and radius for a galaxy that has lost various percentages of its dark and luminous mass due to tidal stripping.  This may provide an order of magnitude estimate for the impact of stripping on WLM, but the models assume a dwarf galaxy with a King profile and minimal rotation, so only a qualitative comparison is possible.  While WLM's location on this diagram is not evidence itself for an isolated history, we can say that if it were to go through significant tidal evolution it would likely transition down into the sequence of the diagram occupied by the majority of the dSphs.

A more appropriate comparison may be to consider WLM with respect to the dwarf elliptical (dE) satellites of M31.  Studies of NGC 147, NGC 185, and NGC 205 have shown them to be of similar total mass and metallicity as WLM \citep{Bender91,deRijcke06,Geha10} although with higher luminosities and gas fractions more than two orders of magnitude smaller \citep{Mateo98}.  Analyses of the velocity fields of NGC 147 and NGC 185 at large radii by \cite{Geha10} suggest that these objects have similar kinematics as WLM, with all three exhibiting well defined rotation.  However given the larger axial ratios of NGC 147 and NGC 185, their $(V/\sigma)^{*}$ values are 0.95 and 0.91 respectively \citep{Geha10} compared to $(V/\sigma)^{*} \simeq 1.1$ for WLM.  Therefore while the dEs are more similar dynamically to WLM than the dSphs, the dEs likely have stronger anisotropy amoung their stars compared to WLM based on this ratio.  Additionally the mass to light ratio (M/L) of WLM is larger than the dEs despite having similar total masses at comparable radii.  Specifically, while the M/L ratios for the three dEs are all consistent with $M/L = 4$ in solar units \citep{deRijcke06}, WLM's ranges from $ 16 \leq M/L \leq 30$ over the half light radius to the last measured point. From our circular velocity curve in Figure \ref{fig:enclmass} we estimate WLM's dynamical mass at two effective radii to be $M(2R_{e}) \simeq 1\times10^{9}$M$_{\odot}$.  This is very near to the mass of NGC 147 and NGC 185 at similar radius ($5.6\times10^{8},7.2\times10^{8}$; \citealt{Geha10}), but slightly lower than NGC 205 ($2\times10^{9}$; \citealt{Geha06}).  While WLM dynamically resembles NGC 147 in particular, it would seem that the relative dark matter content of WLM is somewhat higher than all of the Local Group dEs.  However it is interesting to note that a conversion of WLM's remaining gas mass to stellar mass would result in a $(M/L) \simeq 4$, consistent with the dEs.

We note that the derived mass for WLM is above the possible threshold for gas retention found in the dwarf galaxy simulations of \cite{Sawala11}.  That paper, along with recent simulation and observation work in the last two years \citep{Governato10,SJ10}, indicate that the global structural and dynamical properties of dwarf galaxies with gas are more strongly impacted by the initial halo mass, and their subsequent evolution driven by internal feedback or secular evolution, rather than tidal effects from a larger host galaxy.  However environment likely still plays some role given that WLM and the dEs (which both appear to have comparable total masses within two effective radii) have drastically different gas fractions and effective radii - presumably due to the fact that the dEs reside within the virial radius ($\leq 200$kpc) of M31 where ram pressure and tidal stripping is expected to be efficient.  Differences such as these make it even more important to characterize the evolution of isolated galaxies, and the analysis in this work suggests that WLM represents one possible outcome of dwarf galaxy evolution in the absence of strong environmental forces.\\
 \\
\section{Discussion}
In the last sections we have presented data on the structure and dynamics of WLM, which shows a thick gaseous and stellar configuration, as well as equal pressure and rotation support in the stellar kinematics.   Constraints based on the current position and velocity of WLM with respect to the Local Group barycentre indicate that WLM could only have had at most one, if any, pericentre passage(s) and spent most of its history in isolation, with minimal tidal interactions in its lifetime.  Thus we might ask how the extended structure of the gas and stellar disks/spheroids, and relatively high pressure support in the stars came about?  As discussed in \cite{Brooks07,Oppenheimer08,SJ10,Governato10,Roychowdhury10,Sawala11}, the effect of stellar feedback (SFR, SNe) is increasingly important in galaxies as their total mass decreases.  Other internal processes that can work to transform the orbits and structure of stellar and gaseous components include global disk instabilities, bars, and giant molecular cloud (GMC) heating \citep{SS51,SS53,Sotnikova03}. These effects may also increase with time in a system with ongoing SF like WLM.  We can gain insight into these processes by studying the change in velocity dispersion over the course of WLM's history.\\
\subsection{Velocity Dispersion Evolution}
\begin{figure}
\begin{center}
\ifpdf
\includegraphics[width=0.98\textwidth]{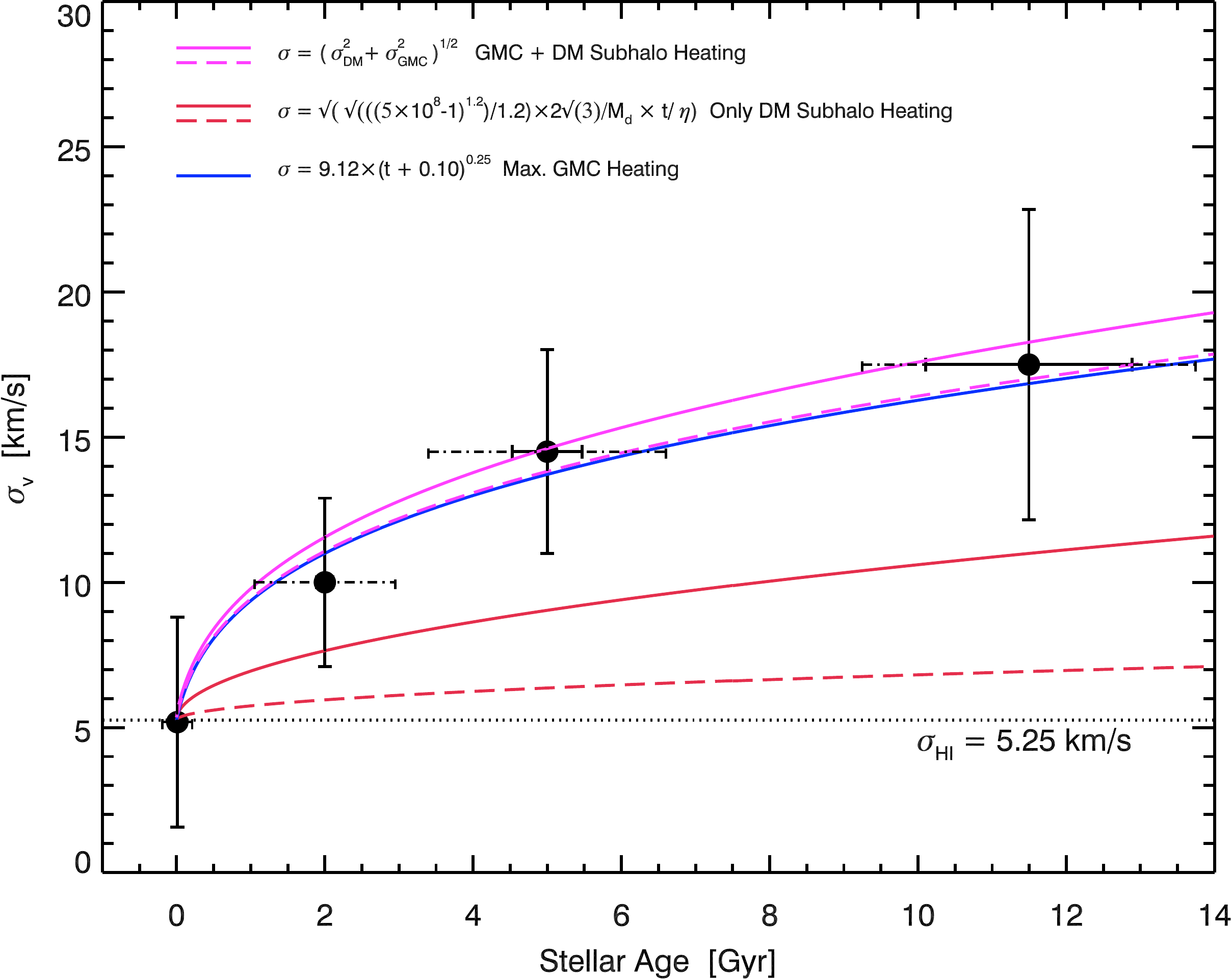}
\else
\includegraphics[angle=270,width=0.98\textwidth]{wlmsigtt2.eps}
\fi
\caption{Velocity dispersion as a function of age in the stellar components of WLM.  Random error on the mean and systematic errors for each age bin are shown as the solid and dash-dot error bars respectively. (Youngest data point ($t = 0.01$) is calculated from the supergiant study of \cite{Bresolin06}.  Blue dashed line indicates the maximum expected heating from simulations of giant molecular cloud heating, and red solid and dashed line represent an upper and lower contribution from DM substructure heating (see text).  Composite curves including both effects added in quadrature is shown in magenta.}
\label{fig:wlmsigt2}
\end{center}
\end{figure}

In Figure \ref{fig:wlmsigt2} we show the velocity dispersion in stellar populations of different ages.  There is a trend of increasing velocity dispersion with increasing stellar age, and at all ages the stellar dispersion is above that of the HI.  It is useful now to discuss how exactly the stellar dispersions may have increased $\sim 10$ km s$^{-1}$ higher than the neutral gas velocity dispersion. 
One interpretation of Figure \ref{fig:wlmsigt2} is that the stars of all ages are heated at some rate, and originated with an initial, constant dispersion - which in our case agrees with the measured HI velocity dispersion from \cite{Kepley07} (dotted line)\footnote{The dispersion of the gas that stars formed out of may be even colder than the HI dispersion if one considers the formation occurring in the cold molecular phase}.   Alternatively, stars of different ages could have formed with different dispersions depending on the ISM kinematics at the time of their birth.  Given that it is possible both interpretations are working in combination, we now briefly discuss a few processes that might be expected to contribute to heating of the stellar populations in WLM.

\subsubsection{Tidal Heating}
\cite{Mayer01b} and \cite{Mayer06} showed that tidal effects of a dwarf galaxy passing through pericentre on its orbit about the Milky Way may trigger bar formation, which later undergoes a bending instability and then works to increase dispersion in the stellar components.  However, this bar formation is not expected to occur (at least to a strong degree) in populations that are already heated, or born thick \citep{Kazantzidis11}.  Recent simulations that use dIrr progenitors modelled with more realistic SF and feedback effects and inject them into the orbits of \cite{Mayer06} seem to show that the structural and dynamical transformation is also possible through direct tidal interactions between the stars of the dwarf galaxy and the potential of the MW \citep{Mayer11}.  This conclusion appears to be at odds with \cite{Penarrubia08}, who showed that dispersion decreases at all radii as tidal stripping progresses - a trend that goes in the opposite direction than what is seen in Figure \ref{fig:wlmsigt2}.   However, if new stars continue to form as the dwarf is tidally stripped of mass (a scenario not included in \cite{Penarrubia08}), perhaps both models could lead to lower velocity dispersion at younger ages.  We also note that even the simulations of \cite{Mayer11} that produce hot spheroids through direct tidal heating only place their dwarfs on orbits that have apocentres of $r_{apo} = 250$ kpc.  The implied apocentre of $r_{apo} \geq 1$ Mpc for WLM (Figure \ref{fig:wlmiso}) and orbital time of $\geq 11$ Gyr,  limits it to having had at most one pericentre passage, and would seem to rule out such a tidal scenario for WLMs heating curve.

\subsubsection{ISM Pressure Floor}
Alternatively, Figure \ref{fig:wlmsigt2} could be interpreted as showing that the oldest stars formed from birth in an originally hotter configuration.  Simulations by \cite{Kaufmann07} found that in lower mass halos, an ISM temperature floor could produce initially thicker dwarf galaxies, as gas pressure support becomes increasingly dominant in lower mass halos.  If this ISM temperature floor decreased with time, perhaps a fraction of the evolution of the stellar velocity dispersion could be attributed to the stars forming out of a hotter ISM at earlier epochs.  In the ``cold flow'' picture \citep{Keres05,Brooks09}, this galaxy should at all times have accreted gas ``cold'' (i.e., on the free fall time), but at high redshift it is likely along filaments, while below $z \sim 1-2$ it would be more spherical accretion.  In this case the filamentary accretion perhaps could drive turbulence in the ISM to a strong degree - which would naturally settle with time as the accretion became more spherical.  Explicit evidence for accretion is difficult to obtain, but the metallicity distribution function for the WLM stars does show a deviation from a closed box chemical evolution model, and in addition the spread in metallicity shows a slight increase as a function of age (Leaman et al., 2012b).

\subsubsection{Giant Molecular Cloud and Dark Matter Subhalo Heating}
The merger history of an object like WLM may also be important in dictating the evolution of its velocity dispersion and vertical thickness with time.  \cite{Abadi03} and \cite{Robertson06} looked at the relative contribution of mergers of varying masses and gas fractions.  The latter study noted the importance of gas rich merging as a way to maintain rotation support in galaxies, while the work of \cite{Abadi03} showed that a hot, thick stellar disk often forms early on, around the time of the last major merger.  However both studies could not include the high resolution stellar feedback effects shown to be important in \cite{Governato10}, and were simulating galaxies that were several orders of magnitude more massive than WLM.  

The case of heating by giant molecular clouds (GMCs) usually assumes that the clouds are occupying a relatively thin region of a galaxy and the encounter velocity is dictated by the peculiar velocity of the stars in a rotating frame.  Therefore $\frac{\partial\sigma^{2}}{\partial{t}} \propto \frac{1}{\sigma}$ (c.f., \citealt{Fuchs01}), and the heating is not particularly efficient as the rate drops as the random velocity component increases.  This is due to the reduced encounter probability, as the stars spend more time outside the GMC ``layer'' as the dispersion increases.  The blue solid line in Figure \ref{fig:wlmsigt2} shows a canonical relation of the form:\\
\begin{equation}
\sigma(t) \propto (t - t_{0})^{\alpha}.
\end{equation}
The maximum heating that is expected analytically and from simulations of giant molecular cloud collisions alone is $\alpha_{GMC} \leq 0.25$ \citep{Lacey84}.  While the mass spectrum of GMCs in WLM is not precisely known, there is observational evidence to suggest the properties are not strongly dependent on host galaxy \citep{Bolatto08}.  This maximum GMC contribution (indicated by the blue line) produces a curve that comes quite close to reproducing the observed stellar dispersion values.  

An additional source of heating may be present due to interactions between dark matter subhalos and the stellar component of WLM (c.f., \citealt{Moore99}).  These perturbations would have higher encounter velocities and expected heating would be more or less isotropic \citep{Fuchs01}.  A simple calculation for the amount of heating that a host galaxy of halo mass $M_{h}$ undergoes from minor mergers with dark matter subhalos of mass $M_{s}$ can be derived assuming that the rate of heating for a subhalo to deposit energy into the \emph{disk} component of the galaxy is:\\
\begin{equation}
\frac{\partial{E_{d}}}{\partial{t}} = \frac{\Delta{E_{d}}}{t_{df}},
\end{equation}
where, $t_{df}$ is the dynamical friction timescale.\\
\begin{equation}
t_{df} = \frac{1.17}{\rm{ln}(M_{h}/M_{s})}\Big(\frac{M_{h}}{M_{s}}\Big)\frac{r_{h}}{V_{c}}.
\end{equation}
For a given interaction, the amount of energy deposited into the halo component relative to the disk component of WLM scales as $\eta \propto 2.3M_{h}/M_{d} \simeq 27$ for WLM \citep{TothOst92}.  Thus the actual energy input into the \emph{disk} will not be a particularly large fraction of the satellite kinetic energy.
Using the expected unevolved subhalo mass function which is shown to be robust to halo size and over a range of redshifts \citep{Giocoli08}:
\begin{equation}
\frac{\partial{n}}{\partial\rm{ln}(M_{s}/M_{h})} = A\Big(\frac{M_{s}}{f M_{h}}\Big)^{-p} {\rm exp}\Big[-\Big(\frac{M_{s}}{f M_{h}}\Big)^{q}\Big],
\end{equation}
\cite{MBW10} show that the total heating rate by dark matter subhalos is:
\begin{equation}
\frac{\partial{E_{d}}}{\partial{t}} \propto \int_{m_{min}}^{m_{max}} n(M_{s}|M_{h})M_{s}^{2}dM_{s},
\end{equation}
where the expected number of subhalos for a given primary halo can be further simplified to $n(M_{s}|M_{h}) \propto M_{s}^{-1.8}$.  In order to equate this heating rate by the dark matter halos to an observable, we make the assumption that the kinetic energy deposited in the disk by a single event is:
\begin{equation}
\Delta{E_{d}} = (1/2)M_{d}\sigma_{z}^{2}(R).
\end{equation}
The total heating rate due to dark matter subhalos from masses $m_{min}$ to $m_{max}$ can then be rewritten as:
\begin{equation}
\frac{\partial\sigma^{2}}{\partial{t}} \propto \frac{2\sqrt(3)M_{s}^{1.2}}{1.2M_{d}}\Big|^{M_{s}=m_{max}}_{M_{s}=m_{min}}.
\end{equation}

Therefore the expected total heating rate by dark matter substructure can be roughly estimated by fitting our velocity dispersion-age data with a function of two variables; the maximum subhalo mass ($m_{max}$) and the WLM disk mass ($M_{d}$).  We assume equipartition of energy in all velocity directions, which has been shown to be appropriate for satellites with circular orbits of various inclinations, and an isotropic Maxwellian distribution of orbits \citep{TothOst92}.  Similarly, assuming an average dispersion within $1r_{e}$ for an age bin is justified, as the scale affected by the energy injection of satellites has been shown to be at least that large \citep{Sellwood98}.  Therefore it is a reasonable assumption for this simple model that disk stars at nearly all radii are heated.  In red we overlay two examples for a heating model based solely on DM interactions.  Both curves have a maximum perturber corresponding to $m_{max} = 5\times10^{8}$ M$_{\odot}$, or $\sim 5\%$ of the total halo mass of WLM.  The dashed line uses a disk mass of $1\times10^{8}$ M$_{\odot}$, close to the present day value, while the solid line assumes a disk mass of $8\times10^{5}$ which is approximately the disk mass of WLM 11 Gyr ago using the scaling relations of \cite{Dutton09}.  Given the assumptions discussed in this paragraph, both curves are strong upper limits on the amount of heating substructure would predict.  Despite this, the fact that the energy deposited in the disk scales as the mass ratio of WLM's disk to halo, prevents strong contribution from this effect.  Even computing substructure models with an unphysical value of $\eta = 1$ does not match the observed dispersion evolution.  This is in good agreement with semi-analytic and n-body simulations of disk heating such as \cite{Benson04}, which found similar inefficiencies for substructure heating.  

In that study the authors showed that for lower mass primary galaxies, the contribution of heating from GMC encounters will dominate relative to substructure interactions.  As an additional independent check, we compute the heating/thickening predicted by Equations 11 and 12 of \cite{Benson04}, expressed as a ratio of vertical to radial scale lengths, $h = z_{x}/r_{d}$.  With modest assumptions for the parameters of those equations we find the same trend - GMC heating can produce a system with $0.32 \leq h \leq 0.57$, close to the observed inclination corrected axial ratio of WLM (see $\S 4.4$).  Using the same range of disk and perturber masses in our Equation 10, we again find that the substructure mechanism is not enough to produce the observed vertical scale height - with only $0.02 \leq h \leq 0.09$ found at the end of 12 Gyr in those models.  However recent simulations have shown that disk heating by dark subhalos may be more effective if higher mass ratio accretion events on plunging radial orbits are considered,  however the probability of such events occurring is still thought to be low (Tjitske Starkenburg, private communication).
While there are obvious degeneracies that prevent us from accurately knowing the exact portion of the heating that came from GMC or DM substructure encounters (especially given the uncertain merger rate as a function of time), combined they are clearly of the right order of magnitude to explain the increased velocity dispersion and thickening in WLM.  Most importantly these processes can occur in a galaxy in isolation, and don't require influence from the MW or M31 to modify a galaxy.

Additional secular processes may work to heat the disk in the vertical direction, including long scale length global disk instabilities, as well as transient spiral or overdensity features (c.f., \citealt{Sotnikova03}).  However, as those features may only last 1 Gyr, we can not comment on their impact directly.  Although we have not ruled out a scenario in which gas dispersion decreases with time ($\S 6.1.2)$, \emph{the dispersion evolution presented in Figure \ref{fig:wlmsigt2}, when coupled with WLM's isolation, strongly suggests that the gaseous and stellar populations became heated and thickened through continual internal effects such as GMC and substructure interactions, rather than tidal transformations triggered by the MW.}\\
\\
\subsection{Comparing WLM to Evolutionary Models}
How do these observations fit in with the larger picture of dwarf galaxy formation and evolution?   The structure and dynamical evolution of this isolated dwarf galaxy can be examined in perhaps one scenario that is consistent with the findings presented in this paper.  

From \cite{Sawala11} we see that the ability of a dwarf galaxy to retain gas until the present day may be due in most part to the total halo mass of the dwarf - with a threshold of approximately $1 \times 10^{9}$ M$_{\odot}$ below which the galaxies are less likely to retain gas.  Gas rich galaxies in that simulation also tended to have a late infall ($z \leq 0.2$) to the group, and pericentre distances that were greater than 300 kpc.  Because of this, dIrrs could have gas over a larger percentage of their history, and experience ongoing SF and SNe - internal feedback that contributes to heating the dwarf, creating a slightly puffier system.  Additionally, simulations by \cite{Schroyen11} show that dwarf galaxies with high angular momentum tend to have SF continuously over the full radial extent of the galaxy - compared to non rotating systems in which the gas can efficiently fall to the center and undergo extreme centralized SF and blowout events.  As discussed in \cite{Robertson08}, if SF is regulated by the ability of the gas to settle into the molecular phase, a thick, hot ISM would naturally result in low efficiency SF, in line with the high gas fraction seen for WLM today.  Therefore higher mass, rotating dwarf galaxies would more likely be the type where we would expect to see signatures of extended internal feedback effects - namely a thick stellar configuration.  These systems, if isolated, will be solely subject to internal feedback effects such as SF, GMC interactions and secular global disk instabilities (c.f. \citealt{Sotnikova03}), which only produce mild transformations in their structure and dynamics.
    
   Thus we could have a scenario where the initial mass dictates the probability of gas retention (i.e., \citealt{Sawala11}), which in combination with angular momentum \citep{Schroyen11,Koleva09}, sets the level of internal feedback driven heating.  In this picture, WLM could be an example of an isolated dwarf irregular at high enough mass and isolation to allow for ongoing SF driven feedback which maintains a dynamically hot stellar structure, low SF rate, and high gas content.

\begin{tiny}\end{tiny}

\section{Summary and Conclusions}
We have presented the results of a VLT and Keck II spectroscopic survey of 180 RGB stars in the isolated dIrr galaxy, WLM.  Broadband optical and infrared photometry and HI radio observations allowed for the study of the structure and dynamics of stellar and gaseous populations in a galaxy at the outskirts of our Local Group - a unique oppourtunity in resolved spectroscopic studies of dwarf galaxies.  

The main results from our analysis are:

\begin{itemize}
\item Given the high inclination, the apparent axial ratios of both the gas and stars indicate that WLM has a thickened structure $\langle q \rangle = 0.39-0.57$.  However the data is not detailed enough to distinguish between a thick disk, or flattened oblate spheroid.  This thickened structure is consistent with observations of statistically large samples of dwarfs \citep{SJ10,Roychowdhury10}.\\
\item The ellipticities and position angles of the HI, young C-stars and evolved RGB stars are nearly identical.  The photometric and kinematic position angles of the gas and stars are also closely aligned ($\Psi_{*} = 1^{\circ}, \Psi_{gas} = 2^{\circ}$).\\
\item We measure a maximum rotation in the stars of $V_{rot} = 17$ km/s, which after correction for asymmetric drift approaches $V_{rot} = 36$ km/s - in excellent agreement with the HI.  The velocity dispersion profile reaches a maximum of $\sigma \sim 17$ km/s and declines with radius, unlike the flat profiles seen in dSphs.  The ratio of rotation to pressure support, $V/\sigma$, is of order unity out to $1.5 r_{e}$, where it slowly rises to 2.5 - in contrast to the HI, which shows $V/\sigma \sim 7$.\\  
\item  The stellar velocities show $V/\sigma$ ratios that decrease with age, and the velocity dispersion grows with age at a rate that can be explained through primarily GMC heating and substructure interactions - with the former dominating.  The evolution of $\sigma(t)$ is therefore consistent with internal feedback mechanisms, rather than external environmental effects such as tides.\\
\item  Using rotationally derived mass estimates we compute a mass at the half light radius ($r_{h} = 1656 \pm 49$ pc) of $M_{half} \sim 4\times10^{8}$ M$_{\odot}$.\\
\item Put together, the observations paint a picture in which WLM has spent the majority of its evolution in isolation, with its velocity dispersion and structural evolution governed primarily by conditions at formation and subsequent internal feedback effects, rather than tidal events. 
\end{itemize} 
  
  This departure from dwarf irregulars as thin rotationally supported cold disks may have implications for the timescales required for tidal transformation scenarios, and for analysis of the statistical properties of dwarf galaxy populations on the group scale.  WLM represents one look at how only the baseline secular processes in a galaxy will modify it over $\sim 12$ Gyr.  Further studies of dIrr galaxies of differing masses and isolation levels will continue to disentangle to what degree internal feedback and environmental evolution play in forming the distinct classes of Local Group dwarf galaxies we see today.

\acknowledgments
We would like to thank the anonymous referee for comments that helped improve this paper, and Dr. Amanda Kepley for generously sharing her HI data. RL acknowledges support from NSERC Discovery Grants to Don VandenBerg and KV.  The authors acknowledge support from the International Space Science Institute, and useful discussions with associated team members.  AB acknowledges support from the Sherman Fairchild Foundation.  RI gratefully acknowledges support from the Agence Nationale de la Recherche though the grant POMMME (ANR 09-BLAN-0228).  The authors would like to thank M. Rejkuba, S. Cote, E. Skillman, J. Pe{\~n}arrubia, and J. Navarro for useful discussions.  FORS2 observations were collected at the ESO, proposal 072.B-0497.  Some photometric data is based on observations obtained at Cerro Tololo Inter-American Observatory a division of the National Optical Astronomy Observatories, which is operated by the Association of Universities for Research in Astronomy, Inc. under cooperative agreement with the National Science Foundation.  This research also uses services or data provided by the NOAO Science Archive.  VLA and the National Radio Astronomy Observatory is a facility of the National Science Foundation operated under cooperative agreement by Associated Universities, Inc.  Additional data presented herein were obtained at the W.M. Keck Observatory, which is operated as a scientific partnership among the California Institute of Technology, the University of California and the National Aeronautics and Space Administration. The Observatory was made possible by the generous financial support of the W.M. Keck Foundation.  The authors wish to recognize and acknowledge the very significant cultural role and reverence that the summit of Mauna Kea has always had within the indigenous Hawaiian community.  We are most fortunate to have the opportunity to conduct observations from this mountain.

\textit{Facilities:} \facility{VLT:Yepun(FORS2)}, \facility{Keck:II(DEIMOS)}, \facility{Blanco(MOSAICII)}
\bibliography{wlmrefst}

\clearpage

\clearpage

\clearpage



\clearpage

\clearpage

\clearpage

\clearpage


\clearpage

\clearpage

\clearpage


\clearpage

\clearpage

\clearpage

\clearpage

\clearpage
\begin{flushleft}
\begin{deluxetable}{lccc}
\tablecolumns{4}
\tablewidth{0pc}
\tabletypesize{\footnotesize}
\tablecaption{WLM Properties}
\tablehead{
\colhead{Quantity} & \colhead{Value} & \colhead{Description} & \colhead{Reference}
}
\startdata
\vspace{1mm}
$(\alpha,\delta)$ & (00 01 58, -15 27 45) & J2000 & \cite{Mateo98}\\
\vspace{1mm}
$(l, b)$ & (75.85, -73.63) & & \cite{Gallouet75}\\
\vspace{1mm}
E(B-V) & $0.035$ (mag.) & Line of Sight Reddening & \cite{Alan05}\\
\vspace{1mm}
$D_{MW}$ & $985 \pm 33$ kpc & Galactocentric Distance & Average\\
\vspace{1mm}
$e$ & 0.45 - 0.6 & Eccentricity & This work\\
\vspace{1mm}
$q_{0}$ & $\sim 0.38-0.57$ & Intrinsic Axial Ratio & This work\\
\vspace{1mm}
$z_{0}$ & $705 \pm 28$ pc & Vertical Scale Length & This work\\
\vspace{1mm}
$r_{D}$ & $987 \pm 30$ pc & Radial Scale Length & This work\\
\vspace{1mm}
$r_{h}$ & $1656 \pm 49$ pc & 2D Projected Half Light Radius & This work\\
\vspace{1mm}
$h_{0}^{MA}$ & $ 1209 \pm 23$ pc & Major Axis Disk Scale Length & This work\\
\vspace{1mm}
$PA_{kin}$ & $178 \pm 1 ^{\circ}$ & Kinematic Position Angle & This work\\
\vspace{1mm}
$PA_{phot}$ & $179 \pm 2 ^{\circ}$ & Photometric Position Angle & This work\\
\vspace{1mm}
$V_{hel}^{HI}$ & $-130$ km s$^{-1}$ & Heliocentric Velocity & \cite{Jackson04}\\
\vspace{1mm}
$(\frac{V_{rot}}{\sigma})^{HI}$ & $\sim 6$ & HI Rotational to Pressure Support & \cite{Kepley07}\\
\vspace{1mm}
$(\frac{V_{rot}}{\sigma})^{\star}$ & $\sim 1$ & Stellar Rotational to Pressure Support & This work\\
\vspace{1mm}
$M_{\rm v}$ & -14.1 (mag.) & Integrated V Band Absolute Mag. & \cite{vandenBergh94}\\
\vspace{1mm}
$M_{HI}$ & $(6.3 \pm 0.3) \times 10^{7} $ M$_{\odot}$ & HI Mass & \cite{Kepley07}\\
\vspace{1mm}
$M_{\star}$ & $1.1 \times 10^{7} $ M$_{\odot}$ & Stellar Mass & \cite{Jackson07a}\\
\vspace{1mm}
$\frac{M_{g}}{M_{b}}$ & 0.86 & Gas Fraction ($M_{g} = M_{HI}/0.733$) & This work\\
\vspace{1mm}
$\frac{M_{b}}{M_{tot}}$ & $0.05 - 0.10$ & Baryon Fraction ($M_{b} = M_{\star} + M_{g}$) & This work\\
\vspace{1mm}
$\frac{M_{tot}}{L_{V}}$ & $17 - 30 $ M$_{\odot}$ L$_{\odot}^{-1}$ & Direct Mass to Light Ratio (at $r_{half}$ to $r_{last}$) & This work\\
\vspace{1mm}
$\Upsilon = \frac{\eta 9 \sigma_{0}^{2}}{2 \pi G I_{0} r_{HB}}$ & $232 \pm 31 $ M$_{\odot} $ L$_{\odot}^{-1}$ & Theoretical Mass to Light Ratio & This work\\
\vspace{1mm}
$[Fe/H]^{RGB}_{phot}$ & $-1.45 \pm 0.2$ dex & Photometric RGB Metallicity & \cite{Minniti97}\\
\vspace{1mm}
$[Fe/H]^{RGB}_{spec}$ & $-1.28 \pm 0.02$ dex & Spectroscopic RGB Metallicity & This work\\
\enddata
\end{deluxetable}
\end{flushleft}
\clearpage

\begin{deluxetable}{cccc} 
\tablecolumns{4} 
\tablewidth{0pc} 
\tablecaption{Repeated Measurements} 
\tablehead{
\colhead{Matched Pair} & \colhead{ID}  & \colhead{$V_{hel}$} & \colhead{[Fe/H]}
}
\startdata
1 & 26940 & -98.000 & -1.66 \\
1 & 26939 & -105.110 & -- \\ 
2 & 27429 & -123.000 & -0.79 \\
2 & 27428 & -118.350 & -1.03 \\
3 & 27926 & -123.000 & -1.11 \\
3 & 27925 & -135.010 & -1.37 \\
4 & 28220 & -145.000 & -0.97 \\
4 & 28219 & -1333.610 & -- \\
5 & 28395 & -139.000 & -1.24 \\
5 & 28394 & -124.940 &  -- \\
6 & 28986 & -112.000 & -0.76 \\
6 & 28985 & -104.620 & -- \\
\enddata
\tablecomments{First entry of each pair is the FORS2 observation, second the DEIMOS.}
\end{deluxetable}

\begin{deluxetable}{lcc}
\tablecolumns{3}
\tablewidth{0pc}
\tabletypesize{\footnotesize}
\tablecaption{Mass Estimates}
\tablehead{
\colhead{Description} & \colhead{Value} & \colhead{Notes}
}
\startdata
\vspace{1mm}
Mass within $r_{h}$ & $2.1 \pm 0.3 \times 10^{8}$ M$_{\odot}$ & Eqn. 1 of \cite{Walker09}\\
\vspace{1mm}
Mass within $r_{h}$ from $V_{c}^{\star}$ & $4.3 \pm 0.3 \times 10^{8}$ M$_{\odot}$ &   \\
\vspace{1mm}
Mass within $r_{last}^{\star}$ from $V_{c}^{\star}$ & $8.5 \pm 0.3 \times 10^{8}$ M$_{\odot}$ &  Last measured stellar bin = 2.9 kpc\\
\vspace{1mm}
Mass within $r_{last}^{HI}$ from approaching side HI SIS Model& $9.4 \pm 0.5 \times 10^{8}$ M$_{\odot}$ & Last measured HI point = 4.5 kpc\\
\vspace{1mm}
Mass within $r_{last}^{HI}$ from receding side HI SIS Model& $2.2 \pm 0.1 \times 10^{9}$ M$_{\odot}$ & Last measured HI point = 4.5 kpc\\
\vspace{1mm}
Mass within $r_{last}^{HI}$ from stellar SIS Model& $1.2 \pm 0.1 \times 10^{9}$ M$_{\odot}$ & Last measured HI point = 4.5 kpc\\
\vspace{1mm}
Mass within $r_{last}^{HI}$ from stellar NFW Model& $1.6 \pm 0.1 \times 10^{9}$ M$_{\odot}$ & Last measured HI point = 4.5 kpc\\
\vspace{1mm}
Mass within $r_{200}$ from stellar SIS model & $2.6 \pm 0.2 \times 10^{10}$ M$_{\odot}$ & $r_{200}(SIS) = 60.5$ kpc\\
\vspace{1mm}
Mass within $r_{200}$ from stellar NFW model & $8.9 \pm 0.8 \times 10^{9}$ M$_{\odot}$ & $r_{200}(NFW) = 42.0$ kpc\\
\enddata
\end{deluxetable}
\clearpage

\clearpage

\end{document}